%% file: main.tex
\documentclass[lettersize,journal]{IEEEtran}
\usepackage{amsmath,amsfonts}
\usepackage{algorithmic}
\usepackage{algorithm}
\usepackage{array}
\usepackage{textcomp}
\usepackage{stfloats}
\usepackage{verbatim}
\usepackage{graphicx}
\usepackage{cite}
 
\usepackage{mathtools}
\usepackage{graphics, theorem, amsfonts, graphicx, amssymb, cite, mathrsfs}
\usepackage{color}
\usepackage{setspace}
\usepackage{multirow}
\usepackage{rotating}
\usepackage{comment}
\usepackage[font=footnotesize]{caption}
\usepackage[affil-it]{authblk}
\usepackage{algorithm}
\usepackage{algorithmic}
\usepackage{enumitem}
\usepackage{cite}
\usepackage{booktabs}
\usepackage{subcaption}

\usepackage{titlesec}

\PassOptionsToPackage{hyphens}{url}\usepackage{hyperref}

\input{mysymbol.sty}

\usepackage{theorem}
\newtheorem{assumption}{\hspace{0pt}\bf Assumption}

\newtheorem{theorem}{\hspace{0pt}\bf Theorem}

\newtheorem{remark}{\hspace{0pt}\bf Remark}

\newtheorem{definition}{\hspace{0pt}\bf Definition}

\date{\today}

\def\E{\mathbb{E}}

\definecolor{forestgreen}{rgb}{0.13, 0.55, 0.13}
\definecolor{Gray}{gray}{0.9}

\usepackage{todonotes}
\begin{document}

\title{Learning Resilient Radio Resource Management Policies with Graph Neural Networks}

\author{Navid~NaderiAlizadeh,
        Mark~Eisen,
        and~Alejandro~Ribeiro
\thanks{This work was supported in part by ARL DCIST CRA under Grant W911NF-17-2-0181, in part by NSF under Grant CCF-2112665 (the AI Institute for Learning-Enabled Optimization at Scale, or TILOS), and in part by NSF under Grant DMS-2031985 (the NSF-Simons Research Collaboration on the Mathematical and Scientific Foundations of Deep Learning, or MoDL).

N. NaderiAlizadeh and A. Ribeiro are with the Department of Electrical and Systems Engineering, University of Pennsylvania, Philadelphia, PA 19104, USA (e-mails: \{nnaderi, aribeiro\}@seas.upenn.edu). M. Eisen is with Intel Labs, Intel Corporation, Hillsboro, OR 97124, USA (e-mail: mark.eisen@intel.com).

This work was presented in part at the 2020 IEEE International Workshop on Signal Processing Advances in Wireless Communications (SPAWC)~\cite{naderializadeh2020wireless} and the 2022 IEEE International Conference on Acoustics, Speech and Signal Processing (ICASSP)~\cite{RRM_ICASSP2022}.}
}



\maketitle

\begin{abstract}
We consider the problems of user selection and power control in wireless interference networks, comprising multiple access points (APs) communicating with a group of user equipment devices (UEs) over a shared wireless medium. To achieve a high aggregate rate, while ensuring fairness across all users, we formulate a \emph{resilient} radio resource management (RRM) policy optimization problem with per-user minimum-capacity constraints that \emph{adapt} to the underlying network conditions via learnable slack variables. We reformulate the problem in the Lagrangian dual domain, and show that we can parameterize the RRM policies using a finite set of parameters, which can be trained alongside the slack and dual variables via an unsupervised primal-dual approach thanks to a provably small duality gap. We use a scalable and permutation-equivariant graph neural network (GNN) architecture to parameterize the RRM policies based on a graph topology derived from the instantaneous channel conditions. Through experimental results, we verify that the minimum-capacity constraints adapt to the underlying network configurations and channel conditions. We further demonstrate that, thanks to such adaptation, our proposed method achieves a superior tradeoff between the average rate and the 5\textsuperscript{th} percentile rate---a metric that quantifies the level of fairness in the resource allocation decisions---as compared to baseline algorithms.
\end{abstract}


\section{Introduction}
As 5G network deployments are underway across the world and research studies have already begun on future 6G technologies, wireless devices and services are becoming more ubiquitous, leading to wireless communication networks that are becoming increasingly complex. These networks will provide connectivity to devices ranging from sensors and cell phones to vehicles, drones, and mixed-reality headsets, shifting the paradigm of how \emph{things} connect together. This will give rise to ultra-dense deployment scenarios, where a massive number of transmissions compete to obtain access to a limited amount of wireless resources.

To deal with these challenges, there has been a plethora of work on the problem of \emph{radio resource management} (RRM), where the goal is to efficiently and optimally allocate the limited time/frequency/spatial resources across the wireless network. The approaches proposed in the literature use a wide variety of techniques in optimization, information, and game theories in order to attack various RRM sub-problems, including power control, link scheduling, cell association, sub-carrier assignment, and beamforming~\cite{madan2010cell, shi2011iteratively, yu2013multicell, wu2013flashlinq, naderializadeh2014itlinq, yi2015itlinq+, song2016game, shen2017fplinq}.

Nevertheless, the aforementioned RRM problems in their most general forms are typically NP-hard, implying that as the network size increases, it becomes more challenging to derive exact optimal solutions to them~\cite{luo2008dynamic,liu2013complexity}. That is why most prior work in the literature devises approximate solutions in various regimes of system parameters. With the success of machine learning, and particularly deep learning, over the past few years, learning-based algorithms have emerged to solve challenging problems in wireless communications, including for resource management~\cite{shen2021ai}. As a prominent example, for the class of power allocation problems, several approaches have been proposed using tools from supervised, unsupervised, self-supervised, and reinforcement learning, as well as meta-learning and graph representation learning~\cite{sun2017learning,eisen2019learning,wang2021unsupervised,naderializadeh2021contrastive,song2021supervise,naderializadeh2021wireless,doshi2021deep,naderializadeh2021resource,nikoloska2021fast,he2021overview,shen2020graph,eisen2020optimal,lee2020graph,zhao2021link,wang2021learning,nikoloska2021black}.

A large portion of the prior work on learning-based RRM have considered \emph{unconstrained} optimization of network-wide objective functions, e.g., sum-rate, without any requirements for fair allocation of resources across the network. More recently, the authors in~\cite{eisen2019learning, eisen2020optimal} considered \emph{robust} formulations of the RRM problem, where arbitrary constraints can be included in the optimization problem, such as per-user minimum-capacity requirements. However, in wireless networks, channel conditions fluctuate from time to time and from topology to topology. Therefore, even for a constant number of transmitters and receivers within a given network area, a fixed and strict minimum-capacity constraint may not be satisfiable for some of the receivers with poor channel conditions and is hard to define a priori.

In this paper, we intend to take one step further, and learn \emph{resilient} RRM policies that can adapt the system requirements in a controlled way if the network conditions are so extreme that the original constraints render the RRM problem infeasible. In particular, we consider the joint RRM problems of power control and user selection in a wireless interference network, where the goal is to maximize a network-wide utility function, while ensuring all users in the network are treated fairly. We introduce a resilient RRM formulation, where the ergodic long-term average rate of each user is forced to be lower-bounded by an \emph{adaptive} minimum-capacity constraint, which is learned via an optimized \emph{slack} variable~\cite{chamon2020resilient, chamon2020counterfactual}.

We reformulate the aforementioned constrained optimization problem in the Lagrangian dual domain, and propose a gradient-based primal-dual algorithm to learn optimal RRM policies and their associated optimal constraint slacks, as well as the dual variables corresponding to each constraint in the original optimization problem. We demonstrate how the search over infinite-dimensional RRM policies can be replaced by optimization over a finite set of parameters that can be used to \emph{parameterize} the RRM polices. Under mild assumptions, we prove that such a parameterization only leads to a negligible duality gap, hence enabling us to use the aforementioned primal-dual approach to iteratively update the RRM policy parameters over the course of training. We use a scalable graph neural network (GNN) architecture to parameterize the primal RRM policies, based on a graph topology induced by the underlying instantaneous channel conditions.

We numerically evaluate the performance of our proposed method on a range of system configurations, and show the superior scalability and transferability of the proposed GNN parameterization as compared to baseline methods. We also show how the resilient formulation of the RRM problem trains the per-user slack variables to adapt to the underlying network topology, increasing in value for users in poor network conditions, hence relaxing their minimum-capacity constraints.


The rest of this paper is organized as follows. In Section~\ref{sec_problem_formulation}, we present the system model and formulate the problem. In Section~\ref{sec:primal_dual}, we describe the Lagrangian dual formulation and the proposed primal-dual framework. In Section~\ref{sec:GNN}, we show how the RRM policies can be parameterized using a shared GNN architecture. In Section~\ref{sec:sim}, we present our experimental results. Finally, we conclude the paper in Section~\ref{sec:conc}.

\section{System Model and Problem Formulation} \label{sec_problem_formulation}
We consider a wireless interference network with a set of $m$ access points, or APs, $\{\mathsf{AP}_i\}_{i=1}^m$ and a set of $n$ user equipment devices, or UEs, $\{\mathsf{UE}_j\}_{j=1}^n$, where the APs intend to communicate with the set of UEs across the network. Before communication begins, an AP-UE association procedure takes place, where each UE gets associated with a unique AP to be served by. We assume that this association procedure effectively partitions the network into a set of $m$ disjoint \emph{cells}. In particular, denoting the set of UEs associated with $\mathsf{AP}_i$ by $\mathcal{R}_i\subseteq\{1,\dots,n\}$, 
we have
\begin{subequations}\label{eq:userpool_partition}
\begin{align}
\mathcal{R}_i &\neq \emptyset, \forall i\in\{1,\dots,m\}, \label{eq:non_empty_Tx_userpool}\\
\mathcal{R}_i \cap \mathcal{R}_j &= \emptyset, \forall (i,j)\in\{1,\dots,m\}^2 \text{ s.t. } i\neq j, \label{eq:disjoint_Tx_userpools}\\
\bigcup_{i=1}^m \mathcal{R}_i &= \{1,\dots,n\}, \label{eq:usepools_cover}
\end{align}
\end{subequations}
where~\eqref{eq:non_empty_Tx_userpool} implies that each AP has at least one associated user,~\eqref{eq:disjoint_Tx_userpools} implies that no user is associated with more than one AP, and~\eqref{eq:usepools_cover} suggests that the sets of associated users to different APs cover all the users across the network. For a given user $\mathsf{UE}_j$, we let $[j]$ denote the index of its unique associated AP.

The channel gain between each access point $\mathsf{AP}_i$ and each user $\mathsf{UE}_j$ in the network is a random variable denoted by $h_{ij}$. We collect all the channel gains across the network in a matrix, denoted by $\bbH \in \ccalH \subseteq \mathbb{C}^{m \times n}$, drawn from an underlying distribution $\mathfrak{D}_{\bbH}$. 
Assuming that all transmissions occur at the same time and on the same frequency band, they will cause interference on each other. Therefore, each AP needs to set its transmit power so as to optimize a global, network-wide objective, such as sum-throughput. Moreover, assuming that each AP can serve a single user at each time step, it also needs to decide on which user to serve from its set of associated users. Given a maximum transmit power of $P_{\max}$, we denote the vector of power allocation variables by $\bbp\in [0, P_{\max}]^m$, whose $i$\textsuperscript{th} component, $p_i$, represents the transmit power allocated to $\mathsf{AP}_i$. We also let $\bbgamma\in\{0,1\}^n$ denote the vector of user selection decisions, whose $j$\textsuperscript{th} component, $\gamma_j$, indicates whether $\mathsf{UE}_j$ has been selected to be served by its associated $\mathsf{AP}_{[j]}$. Then, the signal-to-interference-plus-noise ratio (SINR) at each user $\mathsf{UE}_j$ can be written as
\vspace{-.07in}
\begin{align}\label{eq:SINR_def}
\mathsf{SINR}_j(\bbH, \bbp, \bbgamma) = \frac{ \gamma_j \left|h_{[j]j}\right|^2 p_{[j]}}{N + \sum_{i=1, ~i\neq [j]}^m |h_{ij}|^2 p_i},
\end{align}
where $N$ denotes the noise variance. The Shannon capacity of the link between $\mathsf{AP}_{[j]}$ and $\mathsf{UE}_j$ is then given by
\begin{align}\label{eq_capacity}
f_j(\bbH, \bbp, \bbgamma) = \log_2(1 + \mathsf{SINR}_j(\bbH, \bbp, \bbgamma)).
\end{align}

\vspace{-.07in}Due to the aforementioned short-term fading phenomenon, channel realizations vary over time, implying that the power allocation variables also need to be modified temporally. This motivates considering an \emph{ergodic average} rate~$x_j$, which is limited by the ergodic Shannon limit~$\E_{\bbH} [f_j(\bbH, \bbp, \bbgamma)]$, to capture the throughput experienced by each user $\mathsf{UE}_j$ over a long period of time assuming that the underlying fading random process is stationary~\cite{liu2001opportunistic, wang2016dynamic, bazerque2008distributed, wang2011resource}. 
This is motivated by prior studies on optimizing the ergodic performance and characterizing the ergodic capacity regions of time-varying wireless networks, which are well-established problems in the queuing and information theory literature~\cite{neely2010stochastic, neely2003dynamic, li2001capacity}. The goal is to determine power allocation and user selection policies, $\bbp(\bbH)$ and $\bbgamma(\bbH)$, that take as input an instantaneous channel realization $\bbH$ and determine the power levels $\bbp(\bbH) = [p_1(\bbH)~\hdots~p_m(\bbH)]^T$ and user selection decisions $\bbgamma(\bbH) = [\gamma_1(\bbH)~\hdots~\gamma_n(\bbH)]^T$, respectively.\footnote{The implementation of the power allocation and user selection policies, $\bbp(\bbH)$ and $\bbgamma(\bbH)$, requires the knowledge of channel state information (CSI) at the transmitter side (i.e., CSIT), where the APs have knowledge of the CSI of i) their signal links to their associated users, ii) their outgoing interference links to other users, and iii) the incoming interference caused by other APs to their associated users. Such a CSIT assumption can be realized using feedback links, where the users periodically measure and feedback their received signal/interference powers to the corresponding APs.}

We formulate the joint power allocation and user selection problem, which we refer to as the \emph{radio resource management} (RRM) problem, as follows, where we seek to maximize a concave utility $\mathcal{U}(\bbx)$, i.e.,
\begin{subequations}\label{eq_param_problem}
\begin{alignat}{2}
    &\max_{\bbp,\gamma,\bbx} &~~& \mathcal {\mathcal{U}}(\bbx),             \\
    &~~~\text{s.t.} && \bbx       \leq  \E_{\bbH} \left[ \bbf(\bbH, \bbp(\bbH), \bbgamma(\bbH)) \right], \label{eq:ergodic_rate_constraint}  \\
    &&& \bbx \geq \bbf_{\min}, \label{eq:min_rate_constraint_orig}\\
    &&& \bbp(\bbH) \in  [0,P_{\max}]^m, \ \bbgamma(\bbH) \in \Gamma_{n,m}^{\ccalR}.\label{eq:power_constraint}%
\end{alignat}
\end{subequations}
In~\eqref{eq:power_constraint}, the user selection constraint set, $\Gamma_{n,m}^{\ccalR}$, is defined as
\begin{align*}
&\Gamma_{n,m}^{\ccalR} \hspace{-.03in} \coloneqq \hspace{-.03in} \left\{ \hspace{-.025in}\bbgamma(\bbH) \in \{0,1\}^n \hspace{-.005in} \middle| \hspace{-.0025in} \sum_{j\in \mathcal{R}_i}  \hspace{-.03in} \gamma_j(\bbH) = 1, \forall i\in\{1,\dots,m\} \hspace{-.025in}\right\}\hspace{-.03in}.
\end{align*}

Observe that in~\eqref{eq:min_rate_constraint_orig}, we specify a constraint on the long-term capacity for the $i$\textsuperscript{th} user to be at least $f_{i, \min}$. These minimum capacity constraints are included so as to avoid allocating all resources to ``cell-center'' users, which experience higher average signal-to-interference ratio (SIR) levels, hence balancing the RRM policies to treat ``cell-center'' and ``cell-edge'' users (i.e., those with low average SIR values) \emph{fairly}. We further constrain the user selection policy in~\eqref{eq:power_constraint} to $\Gamma_{n,m}^{\ccalR}$, where each AP is only allowed to serve one of its associated users at each time step. 
Note that the formulation in~\eqref{eq_param_problem} allows for an access point $\mathsf{AP}_i$ to not serve any of its associated users at a given time step by setting its transmit power to zero, regardless of the user selection decisions $\{\gamma_j(\bbH)\}_{j\in\mathcal{R}_i}$.

\subsection{Resilient Radio Resource Management}\label{sec:CF}
A fundamental challenge exists in tackling the RRM problem in \eqref{eq_param_problem} due to the potentially unknown or ill-defined minimum-capacity constraints in~\eqref{eq:min_rate_constraint_orig}. Indeed, solving \eqref{eq_param_problem} directly requires explicit a priori knowledge of the minimum-rate requirements, $\bbf_{\min}$. However, such requirements may not be known in practice. Even if these requirements are specified, e.g., by a certain application, they may be infeasible in certain network configurations, i.e., outside the network's information-theoretic ergodic capacity region, due to the complex interference patterns between concurrent transmissions.

We address this problem by introducing 
a \emph{slack} term $\bbz$ for the constraints, and instead find the optimal RRM policies under the loosened constraints~\cite{chamon2020counterfactual, chamon2020resilient}. If for a given user, the original minimum capacity of $f_{i, \min}$ is too strict and not achievable due to poor signal and/or strong interference levels, the additional slack $z_i$ will address such infeasibility by making the constraint adapt to network conditions. However, any increase in slack $z_i$ will render a solution further from the intended solution of \eqref{eq_param_problem}, since an arbitrarily large slack will render the corresponding constraint irrelevant. We, therefore, impose an additional cost on the slack vector $\bbz$, resulting in the \emph{resilient} formulation of \eqref{eq_param_problem}, defined as
\begin{subequations}\label{eq_slack_problem}
\begin{alignat}{2}
    &\max_{\bbp,\bbgamma,\bbx,\bbz} &\; & \mathcal {\mathcal{U}}(\bbx) - \frac{\alpha}{2}\| \bbz \|_2^2, \label{eq:objective_slack}             \\
    &~~~~\text{s.t.} && \bbx    \leq  \E_{\bbH} \left[ \bbf(\bbH, \bbp(\bbH), \bbgamma(\bbH)) \right],  \  \label{eq:rate_constraint}  \\
    &&& \bbx \geq \bbf_{\min} - \bbz,  \label{eq:min_rate_constraint_slack}\\
    &&& \bbp(\bbH) \in  [0,P_{\max}]^m, \bbgamma(\bbH) \in \Gamma_{n,m}^{\ccalR}, \bbz \geq \bb0\label{eq:resource_constraint_slack}.%
\end{alignat}
\end{subequations}
We denote the optimal value of~\eqref{eq_slack_problem} by $P^*$. In \eqref{eq_slack_problem}, along with optimizing the RRM policies $\bbp$ and $\bbgamma$ and ergodic average rates $\bbx$, we also optimize the value of the slack $\bbz$ that optimally trades off the additional utility obtained from relaxing the constraint and the additional cost from the slack itself weighted by $\alpha \geq 0$. The resilient problem is necessarily feasible as $\bbz$ can always be made large to render \eqref{eq:min_rate_constraint_slack} satisfied.  Indeed, the optimal slack $\bbz$ \emph{must} be at least as large as the difference between the user-selected goals $\bbf_{\min}$ and the fundamental minimum rates achievable under the given network configuration. By imposing a negative utility on large values of $\bbz$, the optimization of slack variables will implicitly loosen \eqref{eq:min_rate_constraint_slack} only enough to maximize a tradeoff between the resulting utility $\mathcal{U}$ and the fairness achieved via stricter minimum-capacity constraints.

\begin{remark}[Fairness Measures]\label{remark:fairness}
\normalfont
Note that using minimum-capacity constraints is only one way of enforcing and quantifying fairness, and in general, one can consider a broader family of fairness measures respecting the fundamental fairness axioms~\cite{fairness}. Moreover, in many practical wireless system settings, users have to satisfy certain minimum-capacity requirements to be admitted into the network. Our proposed resilient formulation can effectively prevent this from happening by sacrificing other users' rates by a small amount, hence contributing to the fairness of the resulting allocation of resources across the network.
\end{remark}

\begin{remark}[Negative Slack Values]\label{remark:neg_slack}
\normalfont
Note that in~\eqref{eq:resource_constraint_slack}, we have constrained the slack values to be non-negative, i.e., $\bbz \geq \bb0$. However, certain network configurations might exist with favorable channel conditions, in which the original minimum-capacity constraints can be readily satisfied. In such cases, the restriction on non-negative slack values can be relaxed to enable negative slacks that can further tighten the constraints and enhance the per-user rates. Note that such higher rates can also be encouraged via a monotonically-increasing utility $\ccalU$ in~\eqref{eq:objective_slack}, such as sum-rate, as we will use later in this paper.
\end{remark}

\section{Proposed Primal-Dual Learning Framework}\label{sec:primal_dual}
\subsection{Lagrangian Dual Formulation}\label{sec:lagrangian}
To address the existence of constraints in \eqref{eq_slack_problem}, we reformulate the resilient problem in the Lagrangian dual domain. Despite the non-convexity of the capacity function in \eqref{eq_capacity} rendering the resilient program non-convex, it is known that under mild conditions on the channel distributions (in particular, the distributions having no point of positive probability, i.e., being non-atomic), the RRM problem exhibits zero duality gap~\cite{ribeiro2012optimal}. We can then proceed with the following reformulation without any loss in optimality.

To derive the Lagrangian dual problem, we first introduce the Lagrangian function, with non-negative dual multiplier functions $\bblambda \in \reals_+^n$ and $\bbmu \in \reals_+^n$ associated with each constraint in \eqref{eq_slack_problem}, as
\begin{align}
   &\ccalL(\bbp,\bbgamma,\bbx,\bbz, \bblambda,\bbmu) \nonumber\\
   &~=  {\mathcal{U}}(\bbx) - \frac{\alpha}{2}\| \bbz\|_2^2    - \bblambda^T\left[\bbx - \E_{\bbH} \left[ \bbf(\bbH, \bbp(\bbH), \bbgamma(\bbH)) \right] \right] \nonumber \\
   &~\quad\, - \bbmu^T \left[ \bbf_{\min} - \bbz - \bbx\right]. \label{eq_lagrangian}
\end{align}
The Lagrangian in \eqref{eq_lagrangian} provides a single, unconstrained objective function, which we can optimize using gradient-based methods. In particular, we seek to maximize over the so-called primal functions $\bbp, \bbgamma, \bbx, \bbz$, while subsequently minimizing over the dual functions $\bblambda, \bbmu$, i.e.,
\begin{align}\label{eq_dual_problem}
D^* \coloneqq \min_{\bblambda, \bbmu} \max_{\bbp, \bbgamma, \bbx, \bbz} \ccalL(\bbp,\bbgamma,\bbx, \bbz,\bblambda,\bbmu).
\end{align}
The dual minimization in~\eqref{eq_dual_problem} is considered over all non-negative valued functions while the primal maximization is considered over functions of forms given in~\eqref{eq:resource_constraint_slack}. Note that the Lagrangian needs to include terms corresponding to the constraints in~\eqref{eq:resource_constraint_slack}. However, alternatively, we can enforce these constraints by choosing proper model output functions (e.g., via sigmoid, softmax, and ReLU functions as we will demonstrate later in the paper), which effectively removes the need for including them explicitly in the Lagrangian and reduces the number of required dual parameters. Due to the aforementioned zero duality gap property of the RRM problem, the solution to \eqref{eq_dual_problem} incurs no loss in optimality relative to the original resilient problem in~\eqref{eq_slack_problem}, i.e., $D^* = P^*$.


While the dual formulation in \eqref{eq_dual_problem} removes the complexity of constraints present in \eqref{eq_slack_problem}, it remains a challenging and often intractable problem to solve in practice. Solving \eqref{eq_dual_problem} requires a saddle-point functional optimization---i.e., such optimal policies need to be defined for each possible state $\bbH$ and, therefore,~\eqref{eq_dual_problem} can be considered as an infinite-dimensional optimization problem. 
We address this issue via a 
\emph{parameterized} dual-based framework for solving the resilient RRM problem in \eqref{eq_slack_problem}, or more specifically, its dual version in \eqref{eq_dual_problem}. 

\subsection{Parameterization of the Primal RRM Policies}\label{sec:param_learning}
We propose to tackle the statistical regression problem in \eqref{eq_dual_problem} via \emph{parameterization}, where we replace the infinite-dimensional functional optimization with optimization over a set of parameters of predetermined form. Recall that, in the given problem, we seek to find optimal RRM policies that, once trained, can be generalized to unseen configurations during system operation. 
We thus replace each of the primal RRM policies $\bbg(\cdot)$ with a respective parameterization $\bbg(\cdot; \bbtheta^{\bbg})$ that is fully specified by a finite-dimensional parameter vector $\bbtheta^{\bbg} \in \reals^{q_{\bbg}}$. With this substitution, we obtain the following parameterized Lagrangian function,
\begin{align}
\ccalL_\theta \hspace{-.03in}\left(\bbtheta^{\bbp},\bbtheta^{\bbgamma},\bbx,\bbz, \bblambda,\bbmu\right) \hspace{-.03in} \coloneqq \hspace{-.03in} \ccalL\hspace{-.025in}\left(\bbp(\cdot; \bbtheta^{\bbp}),\bbgamma(\cdot; \bbtheta^{\bbgamma}),\bbx,\bbz, \bblambda,\bbmu\right)\hspace{-.03in},\label{eq_param_lagrangian}
\end{align}
where we have used as inputs to the standard Lagrangian in \eqref{eq_lagrangian} the parameterized RRM policy definitions. 
The parameterized dual resilient problem is subsequently defined as
\begin{align}\label{eq_dual_param_problem}
D_{\bbtheta}^* \coloneqq \min_{\bblambda, \bbmu} \max_{\bbtheta^{\bbp},\bbtheta^{\bbgamma}, \bbx, \bbz}
\ccalL_\theta \left(\bbtheta^{\bbp},\bbtheta^{\bbgamma},\bbx,\bbz, \bblambda,\bbmu\right).
\end{align}

Unlike the unparameterized dual problem in \eqref{eq_dual_problem}, the \emph{parameterized} dual problem in \eqref{eq_dual_param_problem} does not exhibit null duality gap due to the non-convexity of the constraint \eqref{eq:rate_constraint}. Thus, its relation to the original resilient problem in \eqref{eq_slack_problem} is not immediately evident. However, a connection can be made between these two problems by considering a particular class of parameterizations that are sufficiently dense in their representational abilities. We make the following definition of so-called \emph{near-universal} parameterizations:
%
\begin{definition}\label{def_universal}
A parameterization $\bbg(\cdot;\bbtheta^{\bbg})$ with $\bbtheta^{\bbg} \in \Theta$ is a near-universal parameterization of degree $\eps>0$ for functions in $\ccalF$ if, for any $\bbf \in \ccalF$, $\exists\bbtheta^{\bbg} \in \Theta$ such that
\begin{equation}\label{eq_def_bound}
	 \E_{\bbH} \left\| \bbf(\bbH) - \bbg(\bbH;\bbtheta^{\bbg}) \right\|_{\infty}
		\leq \epsilon.
\end{equation}
\end{definition}
Using Definition \ref{def_universal}, we may in fact bound the difference between the optimal value obtained via \eqref{eq_dual_param_problem} and the optimal value obtained via \eqref{eq_slack_problem} despite the non-convexities present in the problem when we utilize near-universal parameterizations to represent the primal RRM policies. In proving this result, we need to introduce some restrictions to the problem formulation that we state as assumptions next.

%
\begin{assumption}\label{assumption_nonatomic}
The probability distribution $\mathfrak{D}_{\bbH}$ is non-atomic in $\ccalH$, i.e., for any set $\ccalE\subseteq\ccalH$ of nonzero probability, there exists a nonzero probability strict subset $\ccalE'\subset\ccalE$ of lower probability, $ 0 < \mbE_\bbH(\ind{\ccalE'}) < \mbE_\bbH(\ind{\ccalE})$. \end{assumption}

%
\begin{assumption}\label{assumption_slater}
Slater's condition holds for~\eqref{eq_slack_problem}. Especially, there exist variables $\bbx_0$, $\bbp_0(\bbH)$, and $\bbgamma_0(\bbH)$ and a strictly positive scalar constant $\sigma>0$ such that 
\begin{align}\label{eqn_assumption_slater}
  \E_{\bbH}\left[\bbf(\bbH, \bbp_0(\bbH), \bbgamma_0(\bbH))\right] - \bbx_0 \geq  \sigma\bbone.
\end{align} \end{assumption}

%
\begin{assumption}\label{assumption_lipschitz}
The expected performance function $\mbE \left[ \bbf(\bbH, \bbp(\bbH), \bbgamma(\bbH)) \right]$ is expectation-wise Lipschitz on $\bbp(\bbH)$ and $\gamma(\bbH)$ for all fading realizations $\bbH$. Specifically, for any pair of power allocation policies $\bbp_1(\bbH), \bbp_2(\bbH) \in [0,P_{\max}]^m$ and user selection policies $\bbgamma_1(\bbH), \bbgamma_2(\bbH) \in \Gamma_{n,m}^{\ccalR}$, there are constants $L_{\bbp}$ and $L_{\bbgamma}$ such that
\begin{align}
  \mbE \| \bbf(\bbH, \bbp_1(\bbH), \bbgamma(\bbH)) &- \bbf(\bbH,  \bbp_2(\bbH),\bbgamma(\bbH)) \|_{\infty} \nonumber \\ 
       &\leq L_{\bbp} \mbE \| \bbp_1(\bbH) - \bbp_2(\bbH)\|_{\infty}, \\
   \mbE \| \bbf(\bbH, \bbp(\bbH),\bbgamma_1(\bbH)) &- \bbf(\bbH,  \bbp(\bbH),\bbgamma_2(\bbH)) \|_{\infty} \nonumber \\ 
       &\leq L_{\bbgamma} \mbE \| \bbgamma_1(\bbH) - \bbgamma_2(\bbH)\|_{\infty}.
\end{align} \end{assumption}

%
Assumptions \ref{assumption_nonatomic}--\ref{assumption_lipschitz} place a set of mild assumptions on the scope of \eqref{eq_slack_problem} and necessary for the subsequent analysis. Assumption \ref{assumption_nonatomic} states that there are no points of strictly positive probability in the distribution $\mathfrak{D}_{\bbH}$, which is a reasonable assumption provided that the channel fading states observed in practice take on a continuum of values. Assumption \ref{assumption_slater} simply states that service demands can be provisioned with some slack, which is generally realizable in \eqref{eq_slack_problem} via sufficiently small $\bbx$ and large $\bbz$. 
Assumption \ref{assumption_lipschitz} is a continuity statement on each of the dimensions of the expectation of the utility function $\ccalU$  and the rate function $\bbf$---we point out this is weaker than general Lipschitz continuity. 

As previously mentioned, the duality gap of the original unparameterized problem in \eqref{eq_slack_problem} is known to be null, implying $D^* = P^*$ \cite{ribeiro2012optimal}. This result does not directly apply to the parameterized dual problem in \eqref{eq_dual_param_problem} due to the restriction of the optimization to a finite-dimensional space of learning parameters. However, given validity of Assumptions \ref{assumption_nonatomic}--\ref{assumption_lipschitz} and using a parameterization that is near-universal in the sense of Definition~\ref{def_universal}, we show that the duality/parameterization gap $|D_{\bbtheta}^*-P^*|$ between problems~\eqref{eq_slack_problem} and~\eqref{eq_dual_param_problem} is small, as we formally state next.


\begin{theorem}\label{theorem_param_duality}
Consider the resilient RRM problem in \eqref{eq_slack_problem} and its parameterized Lagrangian dual in \eqref{eq_dual_param_problem}, in which each of the parameterizations $\bbg(\cdot; \bbtheta^{\bbg})$ is near-universal with degree $\eps_{\bbg}$ in the sense of Definition \ref{def_universal} for both primal RRM policies $\bbg \in \{ \bbp, \bbgamma\}$. If Assumptions \ref{assumption_nonatomic}--\ref{assumption_lipschitz} hold, then the dual value~$D^*_{\bbtheta}$ is bounded as
\begin{equation}\label{eq_theorem_param_duality}
  P^* - \eps_p L_{f} \left\| \tblambda^* \right\|_1
     \  \leq\  D^*_{\bbtheta} 
     \  \leq\  P^*,
\end{equation}
where $\tblambda^* = [\bblambda^*; \bbmu^*]$ denotes the optimal dual variables in \eqref{eq_dual_problem}, $\eps_p =  \max\{ \eps_{\bbp}, \eps_{\bbgamma}\}$, and $L_f = \max\{L_{\bbp}, L_{\bbgamma}\}$.
\end{theorem}

\begin{myproof} See Appendix~\ref{appx:proof_of_duality_gap}. \end{myproof}

In Theorem \ref{theorem_param_duality}, we establish that the solution found via the parameterized dual problem in \eqref{eq_dual_param_problem} is close to that of the original unparameterized resilient RRM problem in \eqref{eq_slack_problem} when near-universal parameterizations are used to represent the primal RRM policies. The degree of this difference is proportional to the degree of near-universality used in the learning parameterization. This allows us to use stochastic primal-dual methods that operate directly on the finite-dimensional, unconstrained problem in \eqref{eq_dual_param_problem}, as we discuss next.

\subsection{Unsupervised Empirical Primal-Dual Learning}\label{sec:empirical_primal_dual}

Since we do not have access to the underlying fading distribution $\mathfrak{D}_{\bbH}$, we resort to an \emph{empirical} formulation of the Lagrangian function~\eqref{eq_param_lagrangian}, where the expectation is replaced by an empirical sample mean. In particular, we draw a sequence of $T$ fading samples $\{ \bbH_{t}\}_{t=1}^{T}$ according to the distribution $\mathfrak{D}_{\bbH}$. This leads to the empirical parameterized Lagrangian
%
\begin{align}
&\hat{\ccalL}_\theta \left(\bbtheta^{\bbp},\bbtheta^{\bbgamma},\bbx,\bbz, \bblambda,\bbmu\right)\nonumber\\
&~= {\mathcal{U}}(\bbx) - \frac{\alpha}{2}\| \bbz\|_2^2  \nonumber \\ 
   &~\quad\, - \bblambda^T\Big[\bbx  - \hat{\E}_{\bbH} \left[ \bbf(\bbH, \bbp(\bbH;\bbtheta^{\bbp}), \bbgamma(\bbH;\bbtheta^{\bbgamma})) \right] \Big]\nonumber \\ 
   &~\quad\, - \bbmu^T \left[ \bbf_{\min} - \bbz - \bbx\right],\label{eq:empirlca_param_lagrangian}
\end{align}
where for any function $\ccalF:\ccalH \to \mathbb{R}$, we define
\begin{align}
\hat{\E}_{\bbH} \left[\ccalF (\bbH)\right] &\coloneqq \frac{1}{T} \sum_{t=1}^T \ccalF (\bbH_t).
\end{align}

\begin{algorithm*}[t]
\setstretch{1.0}
\caption{Primal-Dual Learning of Resilient RRM Policies}\label{alg:learning}
\begin{algorithmic}[1]
\STATE \textbf{Input:} Primal and dual learning rates $(\eta_{\bbp}, \eta_{\bbgamma}, \eta_{\bbx}, \eta_{\bbz}, \eta_{\bblambda}, \eta_{\bbmu})$, \# time steps per configuration $T$.
\STATE Initialize RRM policy parameters $(\bbtheta^{\bbp}_0, \bbtheta^{\bbgamma}_0)$.
\STATE Initialize primal and dual variables $({\bbx}_{0}, {\bbz}_{0}, {\bblambda}_{0}, {\bbmu}_{0})=(\mathbf{0}, \mathbf{0}, \mathbf{0}, \mathbf{0})$.
\STATE Draw a sequence of $T$ fading states $\{ \bbH_{t}\}_{t=1}^{T}$ according to $\mathfrak{D}_{\bbH}$.
\STATE $k \gets 0$.
\WHILE{not converged}
     \STATE Update primal RRM policy parameters (\text{see~}\eqref{eq_pd_update1}-\eqref{eq_pd_update1.5}).
	\STATE Update ergodic long-term average rate and slack variables (\text{see~}\eqref{eq_pd_update2}-\eqref{eq_pd_update3}).
    \STATE Update dual policy parameters (\text{see~}\eqref{eq_pd_update4}-\eqref{eq_pd_update5}).
    \STATE $k \gets k+1$.
\ENDWHILE
\STATE $\left({\bbtheta^*}^{\bbp}, {\bbtheta^*}^{\bbgamma}, {\bbx}^*, {\bbz}^*, {\bblambda}^*, {\bbmu}^*\right) \gets \left(\bbtheta^{\bbp}_k, \bbtheta^{\bbgamma}_k, {\bbx}_{k}, {\bbz}_{k}, {\bblambda}_{k}, {\bbmu}_{k}\right)$.
\STATE \textbf{Return:} Final primal and dual policy parameters and variables $\left({\bbtheta^*}^{\bbp}, {\bbtheta^*}^{\bbgamma}, {\bbx}^*, {\bbz}^*, {\bblambda}^*, {\bbmu}^*\right)$.
\end{algorithmic}
\end{algorithm*}

We can now derive the updates over an iteration index $k$ for each of the primal and dual parameters/variables by either adding or subtracting the partial gradient of $\hat{\ccalL}_\theta \left(\bbtheta^{\bbp},\bbtheta^{\bbgamma},\bbx,\bbz, \bblambda,\bbmu\right)$ with respect to that variable. For power allocation and user selection policies, this gives us the updates,
\begin{align}
\bbtheta^{\bbp}_{k+1} &= \bbtheta^{\bbp}_k  + \eta_{\bbp} \nabla_{\bbtheta^{\bbp}} \left\{ \bblambda^T \hat{\E}_{\bbH} \left[ \bbf(\bbH, \bbp(\bbH), \bbgamma(\bbH)) \right] \right\}, \label{eq_pd_update1}\\
\bbtheta^{\bbgamma}_{k+1} &= \bbtheta^{\bbgamma}_k + \eta_{\bbgamma} \nabla_{\bbtheta^{\bbgamma}} \left\{ \bblambda^T \hat{\E}_{\bbH} \left[ \bbf(\bbH, \bbp(\bbH), \bbgamma(\bbH)) \right] \right\}, \label{eq_pd_update1.5}
\end{align}
where $\eta_{\bbp}, \eta_{\bbgamma} >0$ denote learning rates corresponding to the primal RRM policy parameters variables $\bbtheta^{\bbp}$ and $\bbtheta^{\bbgamma}$, respectively. Moreover, the ergodic average rate and slack variables are, respectively, updated as
\begin{align}
{\bbx}_{k+1} &= {\bbx}_{k}  +  \eta_{\bbx} \left(\nabla_{{\bbx_{k}}}\left\{\mathcal{U}(\bbx_{k})\right\} + \bbmu_{k} - \bblambda_{k}\right), \label{eq_pd_update2}\\
{\bbz}_{k+1} &= \left[{\bbz}_{k}  +  \eta_{\bbz} \left(\bbmu_{k}  - \alpha  \bbz_{k}\right)\right]_+,
\label{eq_pd_update3}
\end{align}
where $\eta_{\bbx},\eta_{\bbz} >0$ respectively denote the learning rates corresponding to the ergodic average rate and slack variables ${\bbx}$ and ${\bbz}$, and 
$[\cdot]_+ \coloneqq \max(\cdot, 0)$. Finally, we descend on the dual variables using the associated partial gradients of the Lagrangian, i.e.,
%
\begin{align}
{\bblambda}_{k+1} &=  \left[{\bblambda}_{k} - \eta_{\bblambda} \left(\bbx_{k} - \hat{\E}_{\bbH} \left[ \bbf(\bbH, \bbp(\bbH), \bbgamma(\bbH)) \right]\right)\right]_+\hspace{-.004in},\label{eq_pd_update4} \\
{\bbmu}_{k+1} &= \left[{\bbmu}_{k} - \eta_{\bbmu} \left(\bbf_{\min} - \bbz_{k} - \bbx_{k}\right)\right]_+,\label{eq_pd_update5}
\end{align}
where $\eta_{\bblambda}, \eta_{\bbmu} >0$ respectively represent learning rates corresponding to the dual variables ${\bblambda}$ and ${\bbmu}$.

The primal-dual gradient updates in \eqref{eq_pd_update1}-\eqref{eq_pd_update5} successively move the primal and dual variables towards the maximum and minimum points of the Lagrangian dual function, respectively. The complete resilient primal-dual learning algorithm is summarized in Algorithm \ref{alg:learning}. Observe that the proposed method is \emph{unsupervised} in the sense that we update the primal, slack, and dual variables so as to optimize the objective function and constraints in~\eqref{eq_slack_problem} directly rather than with labeled solutions.


\begin{remark}[Generalization Across Configurations]\label{remark:family_configs}
\normalfont It is important to note that the original and resilient problem formulations in Section~\ref{sec_problem_formulation}, alongside the primal-dual learning method in Algorithm~\ref{alg:learning}, train the parameterized RRM policies to operate only on a \emph{single network configuration}, i.e., realization of the placement of APs and UEs, the corresponding long-term fading state, and the user-AP association topology. However, in practice, we need the learned RRM policy parameterizations to \emph{generalize} to novel scenarios, unseen during the training process. Therefore, the problem we are actually interested in is that of finding RRM policies that optimize the performance of a \emph{family of configurations}, so that the trained policies can generalize across configurations once training is complete. We defer the details of training resilient RRM policies over a family of configurations, including the corresponding problem formulation and practical considerations,  to Appendix~\ref{appx:family_configs}.
\end{remark}

\begin{remark}[Structured Parameterizations]\label{remark_structured_param}
\normalfont 
Observe that, in the formulation of the parameterized dual problem in \eqref{eq_dual_param_problem} and the subsequent primal-dual algorithm, we make an implicit assumption that the chosen parameterizations properly adhere to the same structure imposed on the unparameterized policies. That is, the primal policy parameterizations take values of associated forms in \eqref{eq:resource_constraint_slack}
. Selecting appropriate parameterizations thus requires consideration of these imposed structures, which can typically be enforced by standard output layer functions (e.g., sigmoid, Softmax, ReLU, etc.)---see Section \ref{sec:GNN_architecture} for details as applied to the RRM problem.
\end{remark}

\section{Parameterizing Resilient RRM Polices via \\Graph Neural Networks}\label{sec:GNN}
The choice of parameterization functions is critical in achieving optimal RRM policies with good practical performance when solving~\eqref{eq_slack_problem}. Fully-connected deep neural networks (DNNs) are a proper choice here, due to their universality property, which states that given enough depth and/or width, they have sufficient expressive power to approximate any function with any desired accuracy~\cite{hornik1989multilayer,eisen2019learning}. However, despite their theoretical properties, such a parameterization does not scale well---as the parameter dimension (particularly in the input and output layers) grows with number of APs and UEs in the network, i.e., $m$ and $n$---and more critically does not generalize over varying network topologies.

In this section, we discuss and develop a graph neural network (GNN) architecture suitable for solving the RRM problem in networks of any size. In particular, we propose to use GNNs as parameterizations for the primal RRM policies outlined in Section~\ref{sec:primal_dual}. 
Broadly speaking, GNNs can be viewed as a generalization of convolutional neural network (CNN) architectures, whose popularity and practical benefits stem largely from their significantly reduced parameter dimension relative to traditional DNNs, their invariance to input size, and their so-called permutation equivariance. GNNs generalize the convolutional operations performed in CNNs with a convolution performed on arbitrarily structured data \cite{bruna2013spectral, henaff2015deep, defferrard2016convolutional, bronstein2017geometric}. Moreover, certain GNN architectures are known to satisfy the near-universality assumption in Definition~\ref{def_universal} for the class of continuous, equivariant functions \cite{keriven2019universal}, thus making them suitable for achieving small error in duality gap. Note that GNN architectures have been previously used in the literature in the context of resource management in wireless network (see, e.g.,~\cite{shen2021ai, eisen2020optimal, naderializadeh2021wireless, shen2020graph}). However, in this work, we specifically leverage GNN-based parameterizations in conjunction with an unsupervised resilient formulation of the RRM problem as outlined in Sections~\ref{sec_problem_formulation}-\ref{sec:primal_dual}, which has not been done in prior work.

\subsection{Graph Construction}\label{sec:graph_construction}
We consider the data structure in the form of a directed graph $\ccalG = (\ccalV, \ccalE, w)$, where $\ccalV$ denotes the set of graph nodes, connected by directed edges in $\ccalE$, and $w:\ccalE \rightarrow \reals$ is a function that determines the edge weights. More specifically, we define $\ccalV=\{1,\dots,n\}$, where each node represents a user. As for the edges, we define $\ccalE$ to include two edge types:
\begin{itemize}[leftmargin=*]
\item \textbf{Signal edges.} These are self-loops from each node to itself that represent the direct link between each user and its serving AP. In particular, for each user $\mathsf{UE}_j$, $j\in\{1,\dots,m\}$, with the associated $\mathsf{AP}_{[j]}$, there is an edge from node $j$ to itself, i.e., $(j,j)\in\ccalE$, with its weight being a function of the channel gain between $\mathsf{AP}_{[j]}$ and $\mathsf{UE}_j$, i.e., $w(j,j)=e(h_{[j]j})$, where $e:\mathbb{C} \to \mathbb{R}$ is an arbitrary function.
\item \textbf{Interference edges.} These are edges representing the interference caused by each AP at its un-associated users. Specifically, for each two distinct users $\mathsf{UE}_j$ and $\mathsf{UE}_{j'}$ with distinct serving APs (i.e., $[j] \neq [j']$), there is an edge from node $j$ to node $j'$, i.e., $(j,j')\in\ccalE$, with its weight being $w(j,j')=e(h_{[j]j'})$.
\end{itemize}

\begin{figure}[t]
\setlength{\belowcaptionskip}{-10pt}
\centering
\includegraphics[page=3, trim = 3.5in 2.5in 4.2in 1.9in, clip, width=0.4\textwidth]{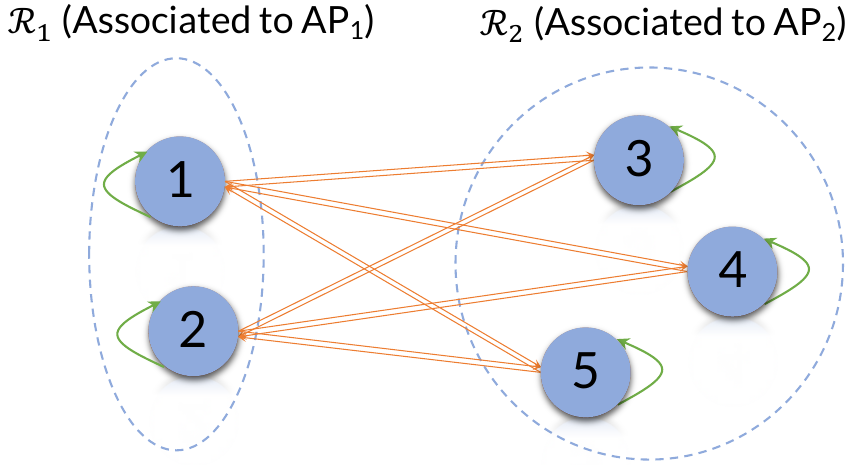}
\caption{The weighted directed graph $\ccalG$ in a network with $m=2$ APs and $n=5$ UEs, where the sets of associated users for the two APs are given by $\ccalR_1=\{1,2\}$ and $\ccalR_2=\{3,4,5\}$. 
The self-loops (in green) represent the \emph{signal edges}, while the edges between each two pairs of connected users (in orange) represent the \emph{interference edges}.}
\label{fig:graph_example}
\end{figure}

Figure~\ref{fig:graph_example} illustrates an example of the graph $\ccalG$ in a network with $m=2$ APs and $n=5$ UEs, where the sets of users served by $\mathsf{AP}_1$ and $\mathsf{AP}_2$ are given by $\ccalR_1=\{1,2\}$ and $\ccalR_2=\{3,4,5\}$, respectively. Based on the graph $\ccalG$, we use a GNN architecture for implementing the primal RRM policies, which we discuss next.

\begin{remark}[Graph Sparsification]\label{remark:sparse_graph}
\normalfont
While it is reasonable to keep all the signal edges in the graph, whether strong or weak, it is also possible to make the graph sparser and structurally asymmetric by removing a subset of interference edges corresponding to ``weak'' interference links in the original network, as for example done in~\cite{naderializadeh2020wireless}. Defining and identifying such ``weak'' interference links is, in and of itself, an interesting research question, and sufficient conditions for information-theoretic optimality of treating interference as noise (e.g., as proposed in~\cite{geng2015optimality, yi2015itlinq+}) can be leveraged to that end.
\end{remark}

\subsection{GNN Architecture}\label{sec:GNN_architecture}
For each node $v\in\ccalV$, let $\bby_v^0\in\reals^{F_0}$ denote the initial \emph{feature vector} corresponding to node $v$, where $F_0$ represents the number of initial features per node. These feature vectors then go through multiple rounds, i.e., \emph{layers}, of message passing and aggregation along the edges of the graph $\ccalG$. Letting $L$ denote the number of such layers, for every layer $l\in\{1,\dots,L\}$, we represent the feature vector of each node $v$ by $\bby_v^l\in\reals^{F_l}$, where $F_l$ denotes the number of features per node at layer $l$. Each such feature vector is derived as an aggregation of the features of node $v$ and its neighbors from the previous layer, as well as its \emph{incoming} edge weights (including its self-loop), i.e.,
\begin{align}\label{eq:GNN_aggregation}
\bby_v^l = \Psi^l\left( \bby_v^{l-1}, w(v,v), \left\{ \bby_u^{l-1}, w(u,v) \right\}_{u\in\ccalN_v}; \bbtheta^l \right),
\end{align}
where $\Psi^l(\cdot;\bbtheta^l)$ denotes a potentially non-linear aggregation function, parameterized through a set of parameters $\bbtheta^l$ , and $\ccalN_v$ is defined as the set of neighbors of node $v$, i.e.,
\begin{align}
\ccalN_v \coloneqq \{u\in\ccalV\setminus\{v\}: (u,v) \in \ccalE\}.
\end{align}

Each aggregation layer in~\eqref{eq:GNN_aggregation} enables each node to receive information about nodes which are one more hop away of itself. It is clear from the proposed graph construction in Section~\ref{sec:graph_construction} that using $L \geq 2$ layers can propagate information from each node to every other node in the graph. After $L$ layers, each node will have a final feature vector $\bbs_v\coloneqq\bby_v^L\in\reals^{F_L}$, which we refer to as its \emph{node embedding}. Note that as the graph edge weights depend on the channel gains, implementing the GNN and deriving the node embeddings requires either iterative message passing among UEs and APs, or having a global entity, which has access to the entire set of channel gains across the network and is able to derive the node embeddings in a centralized manner.

We use the aforementioned architecture as the parameterization for the primal power control and user selection policies. In particular, 
we use a \emph{single} GNN as a \emph{backbone}, i.e., feature extractor, which derives the node embeddings that are subsequently used by \emph{both} the power control and user selection policies in parallel. Such embeddings have been shown to provide semantically-meaningful information when trained and evaluated on wireless power allocation problems~\cite{lee2020graph, naderializadeh2021contrastive}. This helps significantly reduce the total number of parameters as opposed to having a separate, independent GNN for each of the two RRM policies. More precisely, after $L$ aggregation layers 
as in~\eqref{eq:GNN_aggregation}, and obtaining the resulting node embeddings for all the nodes in the graph, i.e., $\{\bbs_v\}_{v\in\ccalV}$, we define the power control and user selection policies as follows:
\begin{itemize}[leftmargin=*]
\item \textbf{Power control.} For each $\mathsf{AP}_i$, $i\in\{1,\dots,m\}$, we derive its transmit power level as
\begin{align}\label{eq:power_allocation_GNN}
p_i(\bbH) = P_{\max} \cdot \sigma\left( \frac{1}{|\ccalR_i|} \bbb_{\bbp}^T \sum_{j\in\ccalR_i} \bbs_j \right),
\end{align}
where $\sigma(\cdot)$ denotes the sigmoid function $\sigma(x)=\frac{1}{1+e^{-x}}$, and $\bbb_{\bbp}\in\reals^{F_{L}}$ is a parameter vector mapping the average node embeddings of the users associated to $\mathsf{AP}_i$ to a scalar, which is then converted to its allocated transmit power. It is evident from~\eqref{eq:power_allocation_GNN} that the resulting allocated power levels satisfy the transmit power constraint $p_i(\bbH)\in[0, P_{\max}]$.

\item \textbf{User selection.} For each $\mathsf{AP}_i$, $i\in\{1,\dots,m\}$ with the set of associated users $\ccalR_i$, 
the \emph{estimated} user selection probability for user $\mathsf{UE}_j, j\in\ccalR_i$ is derived as
\begin{align}
{\gamma}_j(\bbH) &= \mathsf{Softmax}_{\ccalR_i}\left(\bbb_{\bbgamma}^T \bbs_{j} / \tau \right) \nonumber \\
&= \frac{\exp\left(\bbb_{\bbgamma}^T \bbs_{j}/\tau\right)}{\sum_{k\in\ccalR_i} \exp\left(\bbb_{\bbgamma}^T \bbs_{k}/\tau\right)},\label{eq:user_selection_GNN}
\end{align}
where $\bbb_{\bbgamma}\in\reals^{F_{L}}$ is a parameter vector and $\tau\in\reals_+$ denotes a \emph{temperature} hyperparameter. Using the Softmax operation in~\eqref{eq:user_selection_GNN} ensures that $\sum_{j\in \mathcal{R}_i} {\gamma}_j(\bbH) = 1$. However, this will lead to \emph{soft} scheduling decisions, as~\eqref{eq:user_selection_GNN} effectively converts the user node embeddings to a scheduling probability distribution over the set of users associated with each AP. One way to satisfy the \emph{hard} scheduling constraint $\gamma_j(\bbH)\in\{0, 1\}$ is to use a small-enough temperature $\tau \rightarrow 0$, which in turn ``cools'' the resulting distribution and reduces its entropy, mimicking an $\arg\max$ operation, i.e.,
\begin{align}\label{eq:user_selection_GNN_argmax}
{\gamma}_j(\bbH) \xrightarrow{\tau \rightarrow 0} \mathbb{I}\left(\bbb_{\bbgamma}^T \bbs_{j} = \max_{k\in\ccalR_i} \bbb_{\bbgamma}^T \bbs_{k} \right),
\end{align}
where $\mathbb{I}(\cdot)$ denotes the indicator function. Such low temperature values, however, might lead to unstable gradients when updating the policy parameters as in~\eqref{eq_pd_update1.5}. In this paper, we use a more stable alternative, which is to treat the values of $\{{\gamma}_j(\bbH)\}_{j\in\ccalR_i}$ in~\eqref{eq:user_selection_GNN} as a categorical user scheduling distribution. Then, we sample a user based on this probability distribution (i.e., each $\mathsf{AP}_i$ selects one user from $\ccalR_i$ to serve at each step, with the probability of selecting $\mathsf{UE}_j, j\in\ccalR_i,$ given by $\gamma_j(\bbH)$ in~\eqref{eq:user_selection_GNN}), leading to the binary/hard user scheduling decisions $\{\hat{\gamma}_j(\bbH)\}_{j\in\ccalR_i}$. This allows us to replace the user selection policy update in~\eqref{eq_pd_update1.5} with a \emph{policy gradient} update~\cite{eisen2019learning},
\begin{align}
\bbtheta^{\bbgamma}_{k+1} &= \bbtheta^{\bbgamma}_k + \eta_{\bbgamma} \hat{\E}_{\bbH}\bigg[ \left(\bblambda^T \hat{\bbf}(\bbH)\right) \nabla_{\bbtheta^{\bbgamma}} \log \pi_{\bbgamma, \hat{\bbgamma}}(\bbH) \bigg], \label{eq_pd_update1.5_PG}
\end{align}
where $\hat{\bbf}(\bbH)=\bbf(\bbH, \bbp(\bbH), \hat{\bbgamma}(\bbH))$ represents the observed performance function, 
and $\pi_{\bbgamma, \hat{\bbgamma}}(\bbH)$ is defined as
\begin{align}
\pi_{\bbgamma, \hat{\bbgamma}}(\bbH) \coloneqq \prod_{i=1}^m \prod_{j\in\ccalR_i} \gamma_j(\bbH)^{\hat{\gamma}_j(\bbH)},
\end{align}
which represents the joint probability of the selected users by all the APs across the network. Such updates form the basis for the $\mathsf{REINFORCE}$ method in the reinforcement learning literature~\cite{williams1992simple, sutton1999policy, sutton2018reinforcement}. We leave other choices for implementing the user selection policy that may also avoid potentially high variances of policy gradients (e.g., an annealing schedule for the temperature hyperparameter, or using the Gumbel-Softmax distribution and the reparameterization trick~\cite{jang2017categorical, maddison2017the}) for future work.

\end{itemize}

\section{Experimental Evaluation}\label{sec:sim}

\subsection{Wireless Network Settings}
We consider wireless networks with $m\in\{4,6,8,10\}$ APs and $n\in\{40,60,80,100\}$ UEs, dropped randomly within a 500m $\times$ 500m square area. We drop the APs and UEs uniformly at random within the network area, and ensure minimum pairwise distances of 35m and 10m for each AP-AP and AP-UE pair, respectively. The long-term channel model consists of a log-normal shadowing component with 7 dB standard deviation, as well as a standard dual-slope path-loss model~\cite{zhang2015downlink,andrews2016we}, which defines the path-loss at distance $d$ as
\begin{align}\label{eq:PL}
\mathsf{PL}(d)=\begin{cases}
K_0 d^{\alpha_1} &\text{if }d \leq d_{bp},\\
K_0 \frac{d^{\alpha_2}}{d_{bp}^{\alpha_2-\alpha_1}} &\text{o.w.}
\end{cases}
\end{align}
In~\eqref{eq:PL}, we set $K_0=39$ dB, $d_{bp}=100$m, $\alpha_1=2$, and $\alpha_2=4$. We also model the short-term Rayleigh fading using the sum of sinusoids (SoS) technique~\cite{li2002simulation} with a pedestrian speed of 1m/s. The bandwidth is set to 10 MHz, the noise power spectral density is assumed to be $-$174 dBm/Hz, and the maximum transmit power is taken to be $P_{\max} = 10$ dBm.

We adopt a max-SINR user association strategy~\cite{dhillon2013load}, where the set of users associated to each $\mathsf{AP}_i$, $i\in\{1,\dots,m\}$, is defined as
\begin{align}\label{eq:max_SINR_association}
\ccalR_i = \left\{j\in\{1,\dots,n\} : i = \arg \max_{i'\in\{1,\dots,m\}} |h_{i'j}^{\ell}|^2 \right\},
\end{align}
where $h_{i'j}^{\ell}$ denotes the long-term channel gain between between $\mathsf{AP}_{i'}$ and user $\mathsf{UE}_{j}$. 
Note that max-SINR association may be suboptimal in certain scenarios, especially in the presence of load-balancing issues across different APs. Learning optimal user-AP association strategies using GNN architectures is, in and of itself, an interesting resource allocation research problem, where the decisions are made over longer periods of time as opposed to the RRM problems considered in this paper. We leave studying such learning-based user-AP association algorithms as future work.

Once user association is complete, each configuration is run for 200 time steps, with each time step representing 1ms. We use the first 100 time steps as a warm-up period to stabilize the user rates, in which all APs use full transmit power $P_{\max}$ and serve their associated users in a round-robin fashion, and we then use the second $T=100$ time steps to train and evaluate the models. Note that given the aggregate time horizon of 200ms that we consider, the slow speed of the users (resulting in a total of 20cm displacement per user), and the large network area, the impact on the user positions and, therefore, the user-AP association is negligible. Therefore, we assume that the user-AP association does not change over the course of the 200 time steps under study for each configuration.

\subsection{Learning Parameters}
In order to implement the primal RRM policies using a GNN parameterization, we use the local extremum operator proposed in~\cite{ranjan2020asap}, where the aggregation layer~\eqref{eq:GNN_aggregation} is given by
\begin{align*}
\mathbf{y}_{v}^l = \mu\left(\mathbf{y}_{v}^{l-1} \bbtheta_1^l + \hspace{-.08in} \sum_{u: (u, v)\in \mathcal{E}} \hspace{-.07in} w(u, v) \left(\mathbf{y}_{v}^{l-1} \bbtheta_2^l - \mathbf{y}_{u}^{l-1} \bbtheta_3^l \right)\right).
\end{align*}
Here, $\bbtheta_1^l$, $\bbtheta_2^l$, and $\bbtheta_3^l$ are learnable parameters, all in $\mathbb{R}^{F_{l-1} \times F_l}$, and $\mu(\cdot)$ represents a LeakyReLU non-linearity (with a negative slope of $10^{-2}$). Note that with the aforementioned GNN parameterization, the sets of parameters for the power control and user selection policies are given by $\bbtheta^{\bbp} = \left(\{\bbtheta_1^l, \bbtheta_2^l, \bbtheta_3^l\}_{l=1}^{L}, \bbb_{\bbp}\right)$ and $\bbtheta^{\bbgamma} = \left(\{\bbtheta_1^l, \bbtheta_2^l, \bbtheta_3^l\}_{l=1}^{L}, \bbb_{\bbgamma}\right)$, respectively. We use $L=2$ hidden layers, each with 64 features, i.e., $F_1=F_2=64$, and we set the temperature hyperparameter for the user selection policy to $\tau=10$.

We use the normalized channel gains in dB to determine the edge weights. In particular, for a given channel matrix $\bbH$ and for a signal/interference element $h_{ij}$ in $\bbH$, we define the edge weight function $e(\cdot)$ as
\begin{align}\label{eq:GNN_edge_weights_normalization}
e(h_{ij}) = \frac{\log\left({P_{\max}}|h_{ij}|^2 / {N}\right)}{\left(\sum_{i'=1}^m \sum_{j'=1}^n \left[\log\left({P_{\max}}|h_{i'j'}|^2 / {N}\right)\right]^2\right)^{1/2}}.
\end{align}

As for the initial node features, we use a scalar feature for each node, i.e., $F_0=1$, and set it to the \emph{proportional-fairness (PF)} ratio of the corresponding user. In particular, at each time step $t$, for each user $\mathsf{UE}_j$, $j\in\{1,\dots,n\}$, we define the initial node feature vector of node $j$ at that time step as $\bby_j^0(t)=[\mathsf{PF}_j(t)]$, where
\begin{align}\label{eq:PF_def}
\mathsf{PF}_j(t) \coloneqq \hat{f}_j(t) / \bar{f}_j(t).
\end{align}
In~\eqref{eq:PF_def}, $\hat{f}_j(t)$ denotes the \emph{estimated} rate of user $\mathsf{UE}_j$ at time step $t$, defined as
\begin{align}\label{eq:estimated_rate}
\hat{f}_j(t) = \log_2\left(1 + \frac{ P_{\max} \left|h_{[j]j}\right|^2 }{N + P_{\max} \sum_{i=1, ~i\neq [j]}^m |h_{ij}|^2 }\right),
\end{align}
and $\bar{f}_j(t)$ denotes the \emph{exponential moving-average rate} of user $\mathsf{UE}_j$ at time step $t$, which is recursively updated as
\begin{align}
\bar{f}_j(t) = (1-\beta) \bar{f}_j(t-1) + \beta f_j(t),
\end{align}
with $f_j(t)$ denoting the actual achieved rate of the user at time step $t$, and $\beta\in[0,1]$ denoting the inverse averaging window length. In our experiments, we set $\beta=0.05$. Resource management based on the PF ratios 
have been proven to lead to fair resource allocation decisions across the network~\cite{viswanath2002opportunistic}.

We utilize a sum-rate network utility function $\mathcal{U}(\bbx) = \sum_{i=1}^n x_i$, set the minimum capacity to $f_{i, \min} = 1$ bps/Hz for all users in all scenarios, and use a value of $\alpha=10^{-2}$ as the slack norm regularization parameter.

\urlstyle{tt}

As mentioned in Remark~\ref{remark:family_configs} and Appendix~\ref{appx:family_configs}, to make the learned policies generalizable, we train the GNN parameters over a \emph{family} of configurations. In particular, during training, we use 256 training configurations and 128 validation configurations. We train the primal and dual parameters/variables over a total of 400 epochs using a batch size of 64. After each training epoch, we evaluate the trained RRM policies on the 128 validation configurations and save the policies that lead to the highest 5\textsuperscript{th} percentile rate. During the evaluation phase, we test the saved policies on a separate family of 128 test configurations. We initialize the primal RRM policy learning rates as $\eta_{\bbp} = \eta_{\bbgamma} = 10^{-3}$ and the rest of the learning rates as $\eta_{\bbx} = \eta_{\bbz} = \eta_{\bblambda} = \eta_{\bbmu} = 1$. Every 50 epochs, we decrease all of the learning rates by $\frac{1}{2}$. We implement the entire training and evaluation procedures using the PyTorch Geometric library~\cite{Fey/Lenssen/2019}.\footnote{Our code is available at \url{https://github.com/navid-naderi/Resilient_RRM_GNN}.}

\begin{figure*}[t!]
    \centering
    \begin{subfigure}[t]{0.5\textwidth}
        \centering
        \includegraphics[width=.915\textwidth]{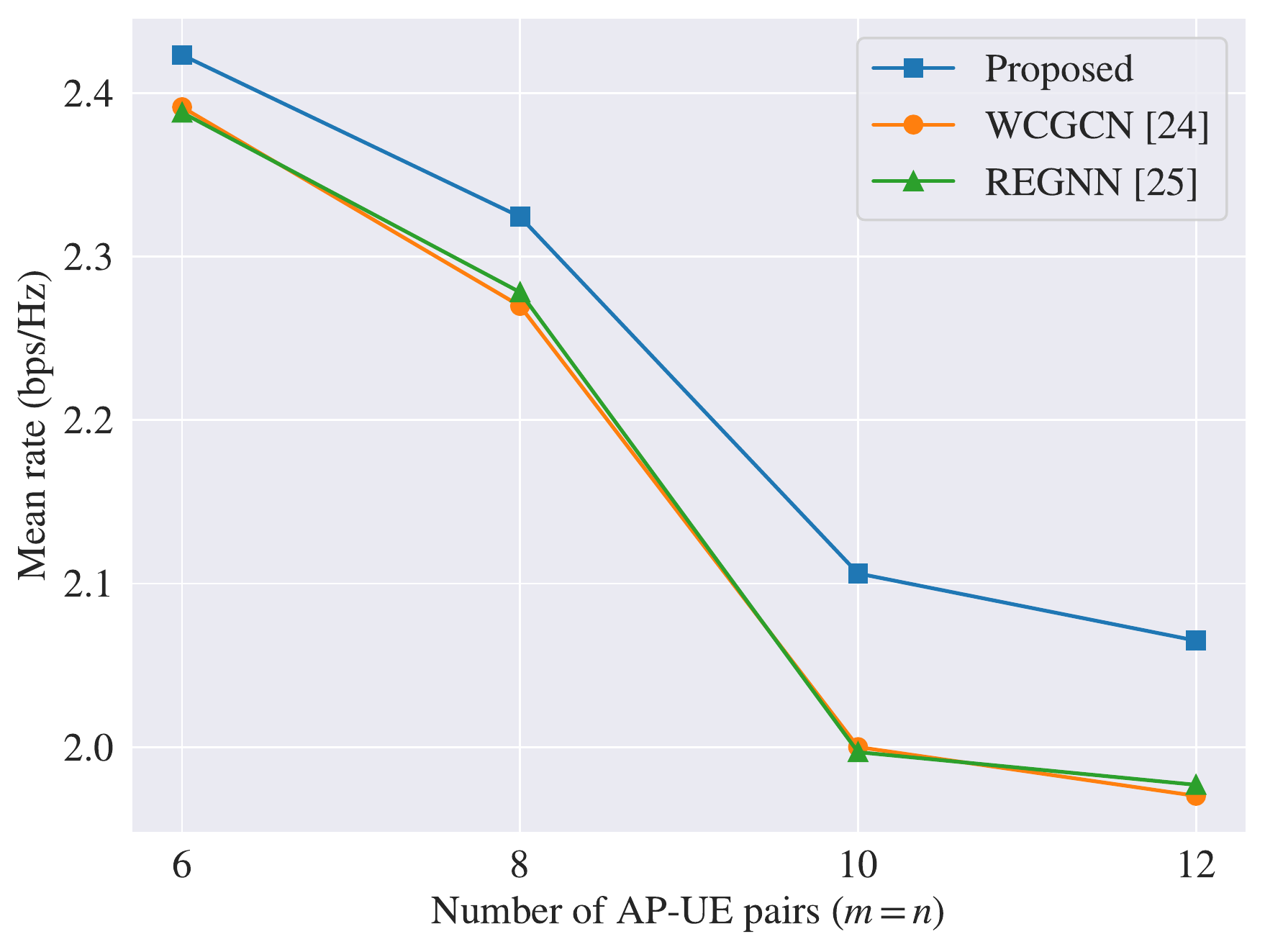}
        \caption{\hspace*{-.4in}}
        \label{fig:mean_same_train_test_n40_powercontrol_GNNbaselines}
    \end{subfigure}%
    \begin{subfigure}[t]{0.5\textwidth}
        \centering
        \includegraphics[width=.95\textwidth]{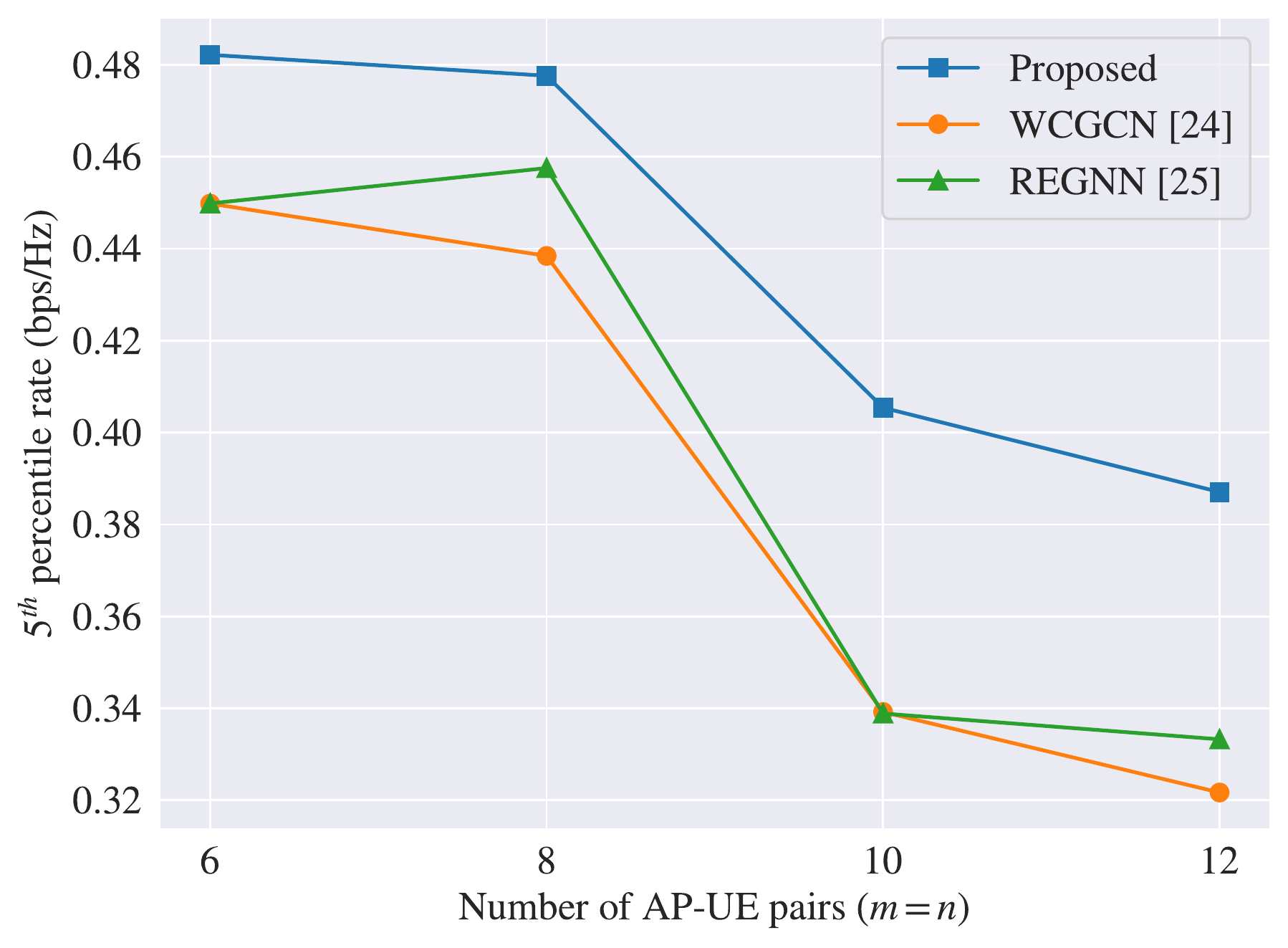}
        \caption{\hspace*{-.45in}}
        \label{fig:rate_5th_percentile_same_train_test_n40_powercontrol_GNNbaselines}
    \end{subfigure}%
    \caption{Power control comparison of the proposed method with GNN-based baselines in terms of (a) mean rate and (b) 5\textsuperscript{th} percentile rate for networks with $m=n\in\{6,8,10,12\}$ AP-UE pairs, where for each scenario, $m$ and $n$ were the same during training and evaluation.}
    \label{fig:same_train_test_n40_powercontrol_GNNbaselines}
\end{figure*}

\subsection{Baseline Methods}
We consider two state-of-the-art unsupervised learning-based baselines using GNN parameterizations, namely WCGCN~\cite{shen2020graph} and REGNN~\cite{eisen2020optimal}, for the problem of power control. We also compare the performance of our proposed method under joint power control and user selection with three non-learning-based baselines, namely full reuse (where all APs transmit with full power at each time step), WMMSE~\cite{shi2011iteratively}, and ITLinQ~\cite{naderializadeh2014itlinq, naderializadeh2017ultra}. 
For the latter baselines, we use a PF-based user selection policy, where at each time step, for each AP, the associated user with the maximum PF ratio is selected.


\makeatletter
\renewcommand\subsubsection{\@startsection{subsubsection}{3}{\z@}%
                                     {-0.3ex\@plus -1ex \@minus -.2ex}%
                                     {-1.5ex \@plus -.2ex}
                                     {\hspace{.12in}\it\normalsize}}
\makeatother

\subsection{Performance on Identical Training and Evaluation Settings}\label{sec:same_train_test_results}
In this section, for the learning-based methods, the evaluation results for any wireless network size (i.e., the number of APs and UEs) are based on the models trained with the same wireless network size.
\subsubsection{Power Control Only ($m=n$)} We first consider networks with equal numbers of APs and UEs, in which user selection decisions are trivial, as only one user is associated to each AP. Therefore, the RRM problem boils down to power control. Figure~\ref{fig:same_train_test_n40_powercontrol_GNNbaselines} shows the performance of our proposed method compared to the GNN-based baseline algorithms for networks with $m=n\in\{6, 8, 10, 12\}$ AP-UE pairs. As the figure shows, thanks to the resilient formulation, our proposed method is able to outperform both GNN-based baselines in terms of the mean and 5\textsuperscript{th} percentile rates.

\begin{figure*}[t!]
    \centering
    \begin{subfigure}[t]{0.5\textwidth}
        \centering
        \includegraphics[width=.93\textwidth]{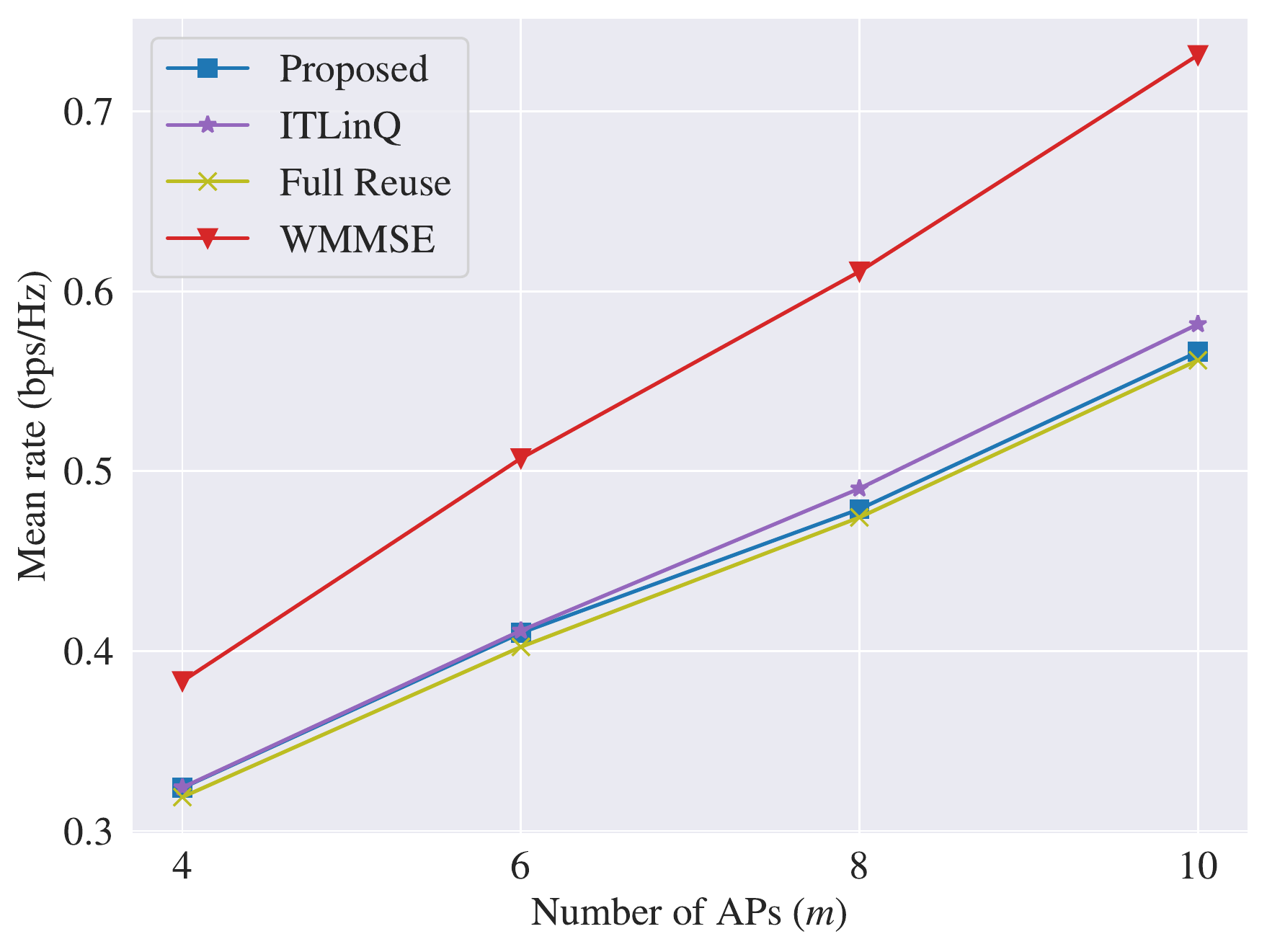}
        \caption{\hspace*{-.4in}}
        \label{fig:mean_same_train_test_n40}
    \end{subfigure}%
    \begin{subfigure}[t]{0.5\textwidth}
        \centering
        \includegraphics[width=.95\textwidth]{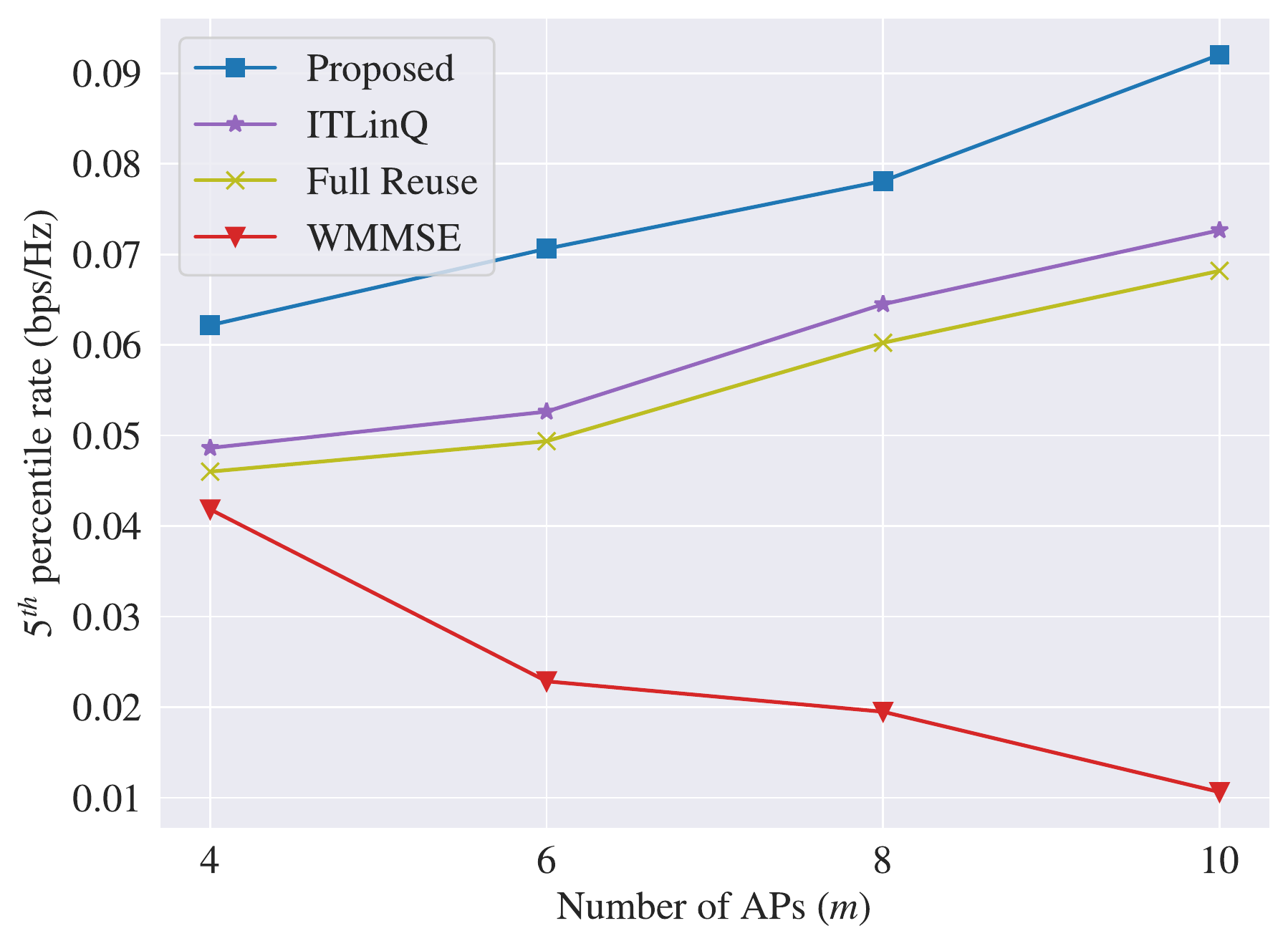}
        \caption{\hspace*{-.45in}}
        \label{fig:rate_5th_percentile_same_train_test_n40}
    \end{subfigure}%
    \caption{Comparison of the proposed method with non-learning-based baselines in terms of (a) mean rate and (b) 5\textsuperscript{th} percentile rate for networks with $m\in\{4,6,8,10\}$ APs and $n=40$ UEs, where for each scenario, the value of $m$ was kept fixed during training and evaluation.}
    \label{fig:same_train_test_n40}
\end{figure*}

\begin{figure}[t]
\setlength{\belowcaptionskip}{-10pt}
\centering
\includegraphics[trim = .1in .1in 0.1in .1in, clip,width=0.42\textwidth]{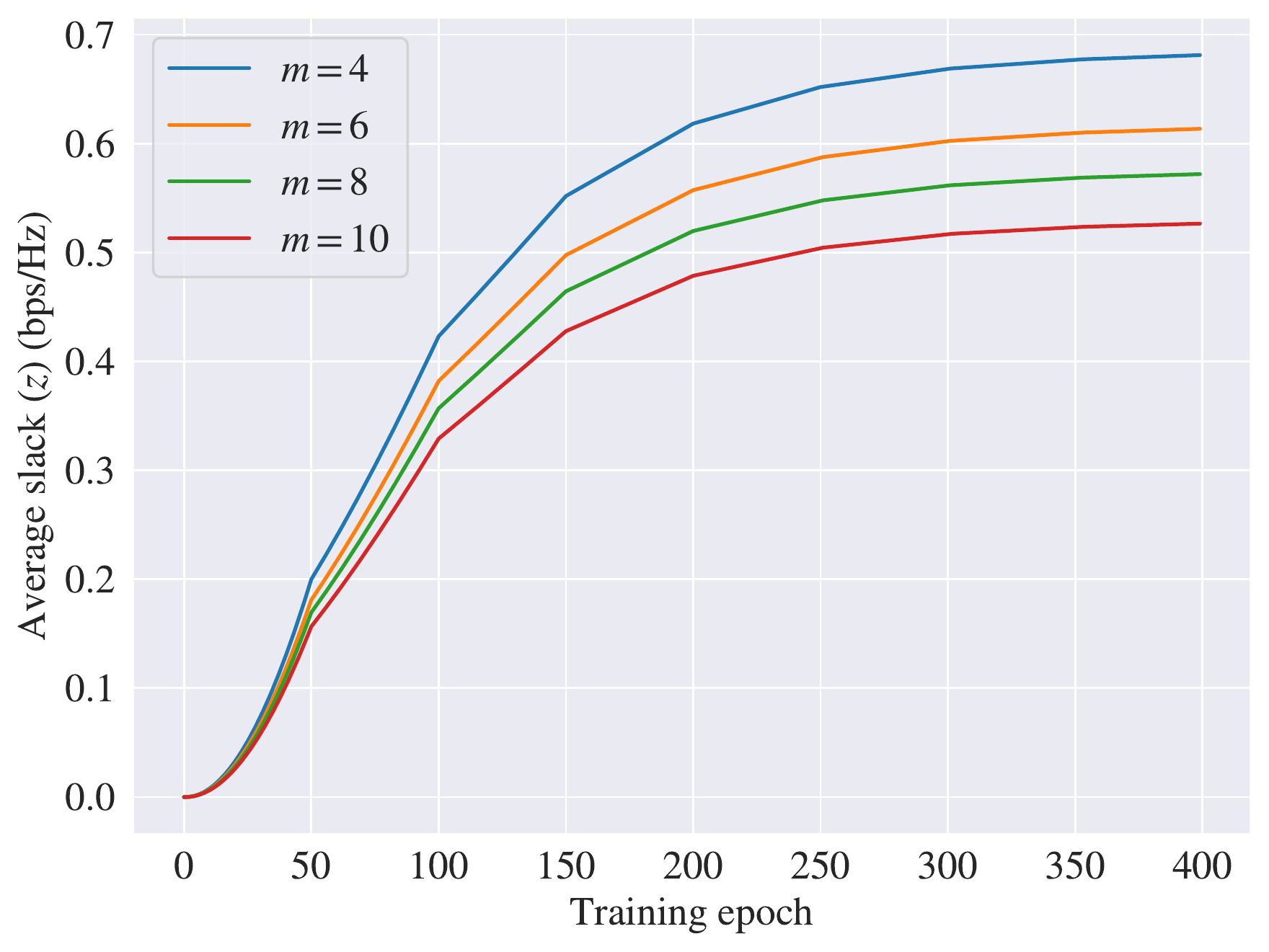}
\caption{Evolution of the average slack variable during training for networks with $m\in\{4,6,8,10\}$ APs and $n=40$ UEs.}
\label{fig:slack}
\end{figure}

\subsubsection{Joint Power Control and User Selection ($m<n$)} Figure~\ref{fig:same_train_test_n40} compares the performance of the proposed method and the non-learning-based baseline methods in terms of mean and 5\textsuperscript{th} percentile rates for networks with $m\in\{4,6,8,10\}$ APs and $n=40$ UEs. As the results show, while underperforming ITLinQ and WMMSE in terms of mean rate, the proposed method significantly outperforms all baselines in terms of the 5\textsuperscript{th} percentile rate. This demonstrates how the resilient formulation of the RRM problem leads to a considerably fairer resource allocation across all the users, balancing the rates achieved by ``cell-center'' and ``cell-edge'' users.

To cast more light on the role of the slack variable in the primal-dual learning process, Figure~\ref{fig:slack} illustrates the evolution of the average slack variable during the training procedure for different values of $m$. As the number of APs, i.e., $m$, increases, each of the $n=40$ users has a higher probability of being served by a closer AP, hence the network will be less interference-limited. This is precisely reflected in Figure~\ref{fig:slack}, where the average slack variable converges to a smaller value for networks with a larger number of APs, hence leading to stricter minimum-capacity requirements.

\begin{figure*}[t!]
    \centering
    \begin{subfigure}[t]{0.5\textwidth}
        \centering
        \includegraphics[width=.93\textwidth]{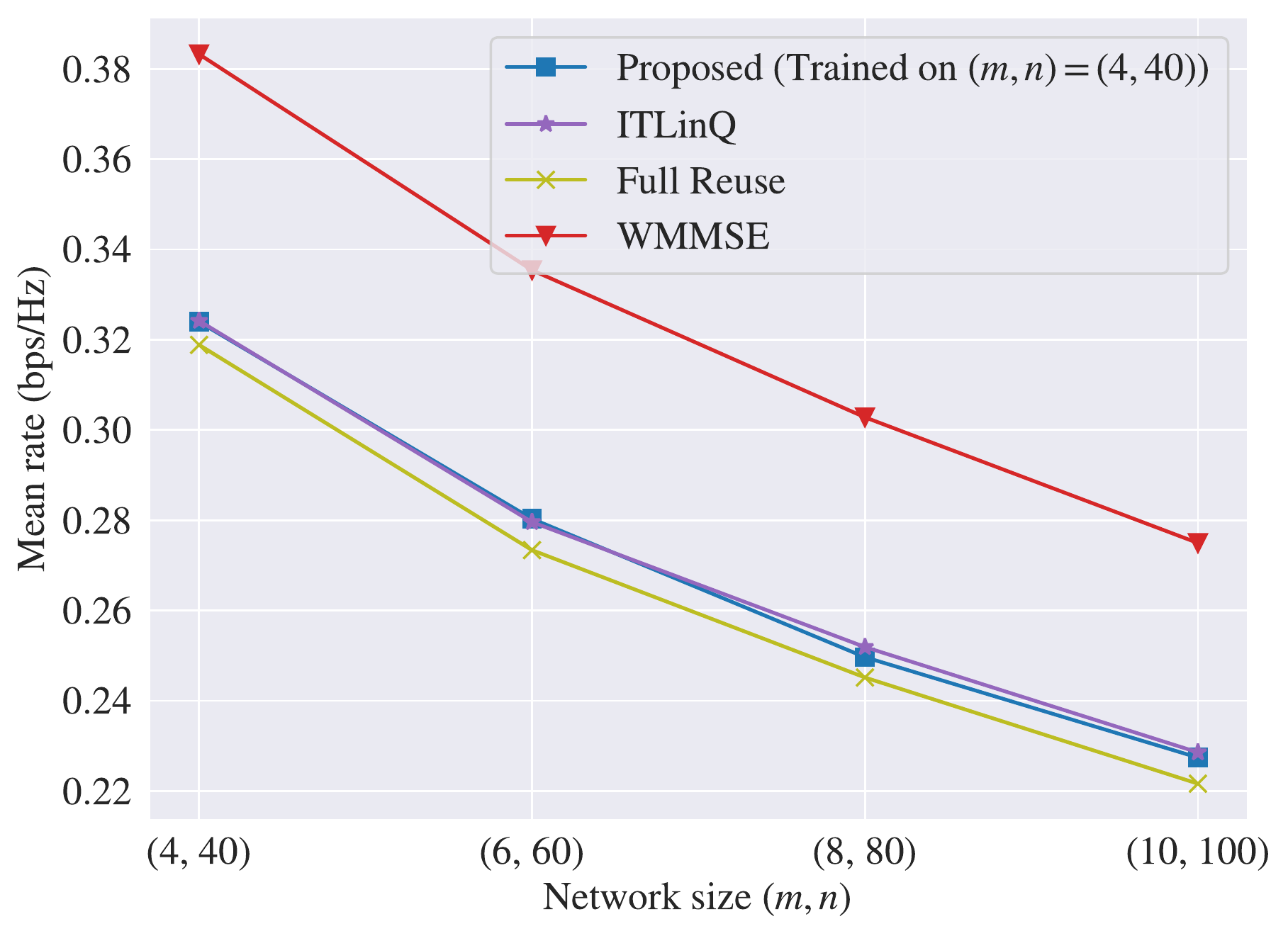}
        \caption{\hspace*{-.4in}}
        \label{fig:mean_transferability}
    \end{subfigure}%
    \begin{subfigure}[t]{0.5\textwidth}
        \centering
        \includegraphics[width=.93\textwidth]{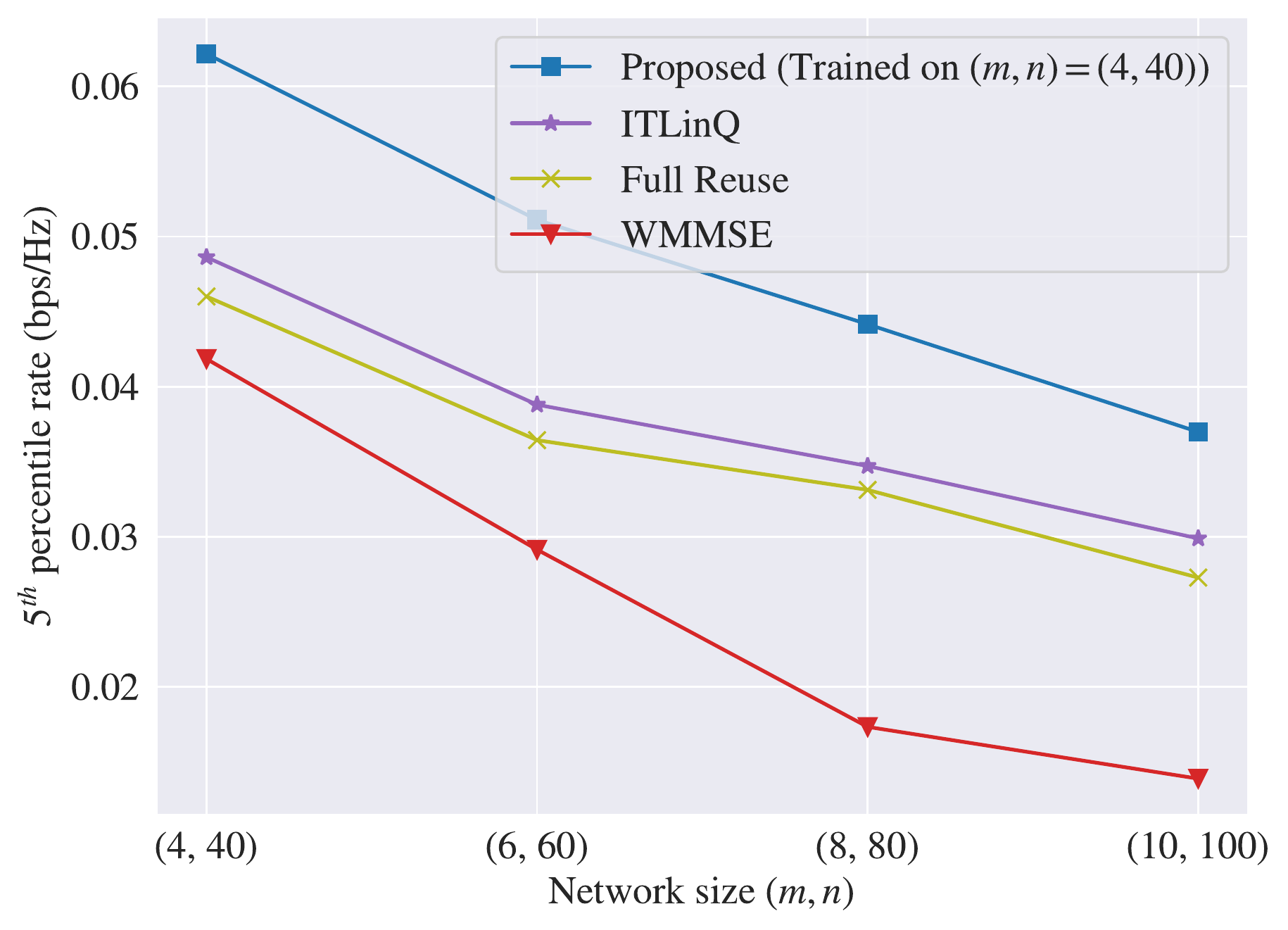}
        \caption{\hspace*{-.45in}}
        \label{fig:rate_5th_percentile_transferability}
    \end{subfigure}%
    \caption{Transferability of the proposed method, where the model trained on networks with $m=4$ APs and $n=40$ UEs is evaluated on larger network configurations. The transferability performance of the proposed method is compared with non-learning-based baselines in terms of (a) mean rate and (b) 5\textsuperscript{th} percentile rate.}
    \label{fig:transferability}
\end{figure*}

\begin{figure*}[t!]
    \centering
    \begin{subfigure}[t]{0.5\textwidth}
        \centering
        \includegraphics[width=.95\textwidth]{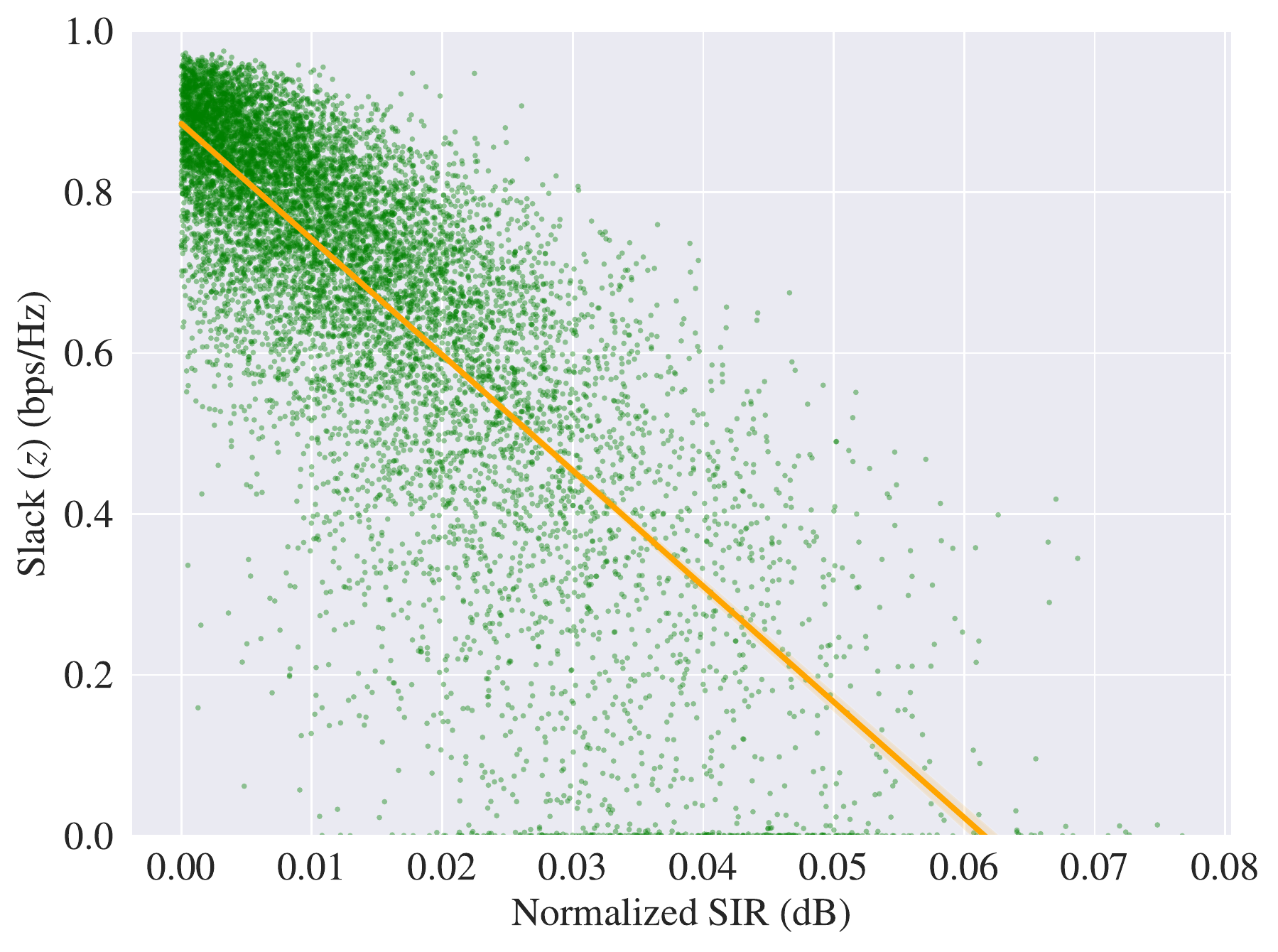}
        \caption{\hspace*{-.25in}}
        \label{fig:scatter_slack_SIR}
    \end{subfigure}%
    \begin{subfigure}[t]{0.5\textwidth}
        \centering\hspace{.3in}
        \includegraphics[width=.89\textwidth]{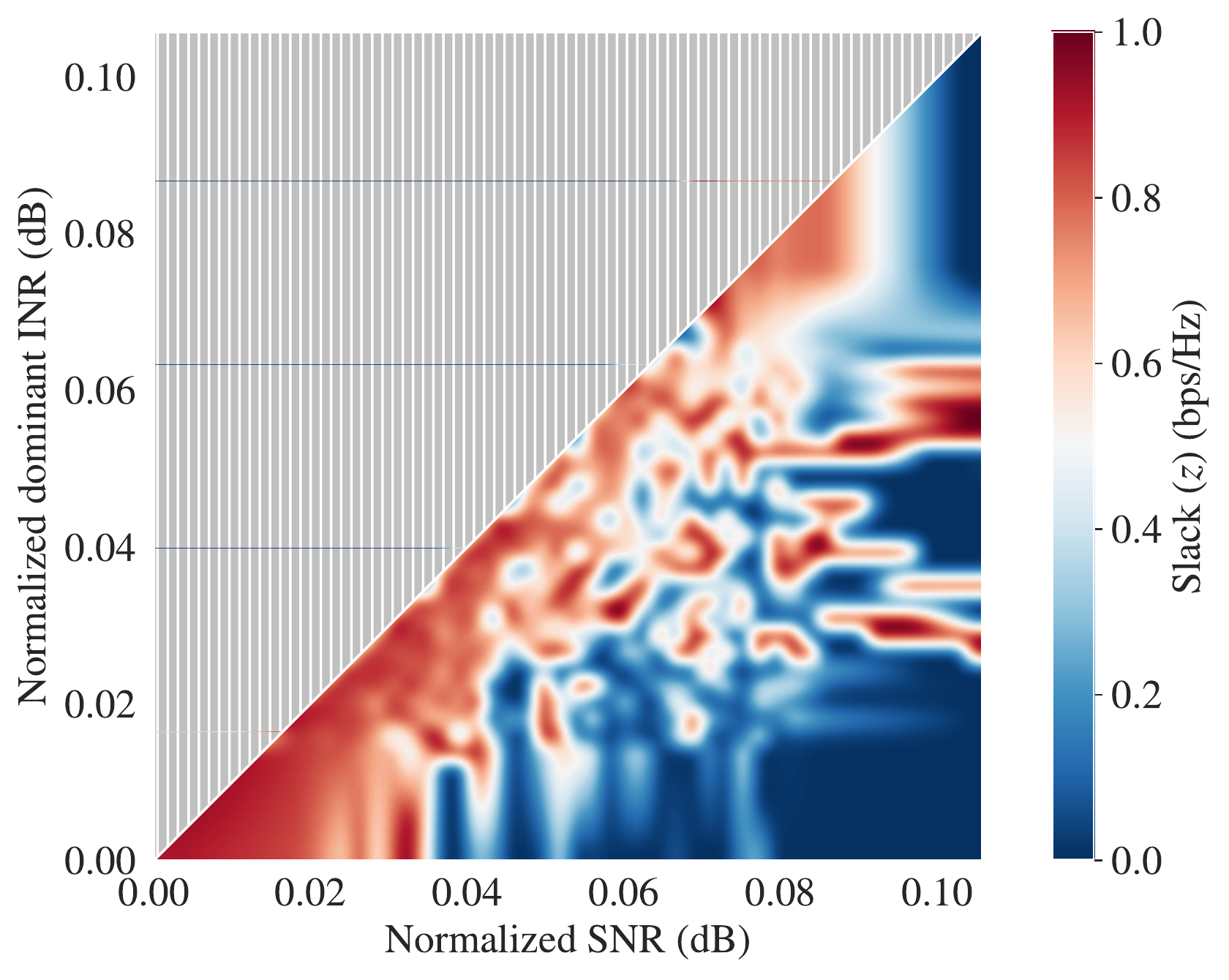}
        \caption{\hspace*{-.2in}}
        \label{fig:heatmap_slack_SNR_INR}
    \end{subfigure}%
    \caption{(a) Scatter plot of the per-user slack values versus the normalized signal-to-interference ratio (SIR) levels, and (b) heatmap of (interpolated) per-user slack values as a function of the normalized signal-to-noise ratio (SNR) and the dominant interference-to-noise ratio (INR). Both plots are based on final slack values for the proposed model trained on networks with $m=4$ APs and $n=40$ UEs. The SIR in (a) and the dominant INR in (b) reflect only the strongest interferer at each user. The SNR, INR, and SIR values are calculated based on the long-term fading states.}
    \label{fig:slack_interpretation}
\end{figure*}

\subsection{Transferability to Larger Network Sizes}
While the results in Section~\ref{sec:same_train_test_results} demonstrated the performance of the proposed method for similar training and evaluation network sizes, as we mentioned in Section~\ref{sec:GNN}, one of the main benefits of GNNs is their \emph{size invariance}. In other words, a GNN trained on a given network size can be evaluated on any arbitrary network size. Thus, here we evaluate the \emph{transferability} of the trained RRM policies, i.e., how policies trained on smaller networks perform in larger configurations.

In Figure~\ref{fig:transferability}, we consider the RRM policies trained on networks with $m=4$ APs and $n=40$ UEs. Once training is complete, we then freeze those policies, i.e., keep their parameters unchanged, and evaluate them on networks with $n\in\{40, 60, 80, 100\}$ UEs and $m=n/10$ APs. As the figure shows, the proposed method transfers significantly well to graphs of more than twice the size, maintaining its 5\textsuperscript{th} percentile rate gains over the baseline methods. This demonstrates the inherent capability of GNN parameterizations that make the resulting models insensitive to the underlying network size, as opposed to regular DNN-based parameterizations, which become unusable if the network size during execution is different from the ones on which the policies have been trained.

\subsection{Interpretation of Slack Values as a Function of Underlying Network Conditions}
Our main motivation for the resilient formulation of the RRM problem was to learn RRM policies that adaptively relax the minimum-capacity constraints for users that are not in desirable channel conditions. To verify that the trained policies have indeed learned such relaxations properly, we can visualize the learned slack variables as a function of network conditions for each user in the training set.

To that end, we consider the primal-dual training procedure in networks with $m=4$ APs, and $n=40$ UEs. For each user $\mathsf{UE}_j, j\in\{1,\dots,n\},$ in each configuration, we retrieve its final slack level $z_j$, alongside three other quantities that reflect its channel conditions, namely:
\begin{itemize}[leftmargin=*]
    \item its large-scale signal-to-noise (SNR), $P_{\max} \left|h^{\ell}_{[j]j}\right|^2 / N$;
    \item its dominant large-scale interference-to-noise ratio (INR), $\max_{i\neq [j]} P_{\max} \left|h^{\ell}_{ij}\right|^2 / N$; and,
    \item its dominant large-scale signal-to-interference ratio (SIR), $\left|h^{\ell}_{[j]j}\right|^2 / \max_{i\neq [j]} \left|h^{\ell}_{ij}\right|^2 $.
\end{itemize}

Figure~\ref{fig:scatter_slack_SIR} shows a scatter plot of the slack values as a function of the large-scale SIR in dB, normalized as in~\eqref{eq:GNN_edge_weights_normalization}. As the figure shows, the slack values have a general downward trend with increased SIR, which is as expected: users with higher large-scale SIR levels have more favorable channel conditions and, therefore, need less relaxation for their corresponding minimum-capacity constraints. Moreover, Figure~\ref{fig:heatmap_slack_SNR_INR} illustrates an interpolated heatmap of the slack value as a function of the normalized SNR and dominant INR levels. The learned slack values are generally largest around the origin (low SNR levels) or the identity line (low SIR levels), which shows how the resilient formulation of the RRM problem provides a granular control over the minimum-capacity requirements for different users across the network.

\section{Concluding Remarks}\label{sec:conc}
We considered the problem of downlink power control and user selection in wireless interference networks with multiple interfering access points (APs), which intend to serve multiple users. To balance fairness across users while maximizing their average achieved rate, we formulated a constrained optimization problem with per-user minimum-capacity requirements. We showed how the aforementioned radio resource management (RRM) policies can be made \emph{resilient} through the introduction of slack variables, which relax the minimum-capacity constraints for users in poor network conditions. We reformulated the problem in the Lagrangian dual domain and introduced parameterizations for the RRM policies to resolve the challenge of infinite-dimensional functional optimization. We specifically used a graph neural network (GNN) parameterization for the RRM policies, and we proposed a primal-dual approach to train the GNN parameters, as well as the remaining primal and dual variables, via iterative stochastic gradient updates. Experimental results demonstrated the superiority of our proposed algorithm compared to baseline methods in terms of the trade-off between average and 5\textsuperscript{th} percentile user rates, even in scenarios where the network size during evaluation was more than twice as large as the ones seen during training. We further showed how the resulting slack variables adapt themselves to the underlying network configuration, increasing in value---thereby relaxing the minimum-capacity constraints---for users with unfavorable channel conditions. 

In this work, we assumed that we have access to the full channel state information across the network to accurately calculate the Shannon capacity values. It would be interesting to study how our proposed method can be implemented in the real world, using noisy and/or quantized values of channel gains, attained through, e.g., periodic channel quality indicator (CQI) feedback. We leave such practical considerations for transferring our method to real-world wireless networks as future work.


\bibliographystyle{IEEEtran}
\bibliography{references}

\appendices
\input{duality_gap.tex}
\input{family_configs.tex}
\vfill

\end{document}

%% file: duality_gap.tex
\section{Proof of Theorem~\ref{theorem_param_duality}}\label{appx:proof_of_duality_gap}

For notational convenience, we begin by restating the Lagrangian in \eqref{eq_param_lagrangian} in a more compact form. To do so, we first define the collected unparameterized primal policies, primal variables, and dual variables as $\tbp(\bbH) \coloneqq [\bbp(\bbH); \bbgamma(\bbH)]$, $\tbx \coloneqq [ \bbx; \bbz]$, and $\tblambda = [\bblambda; \bbmu]$. Further define $\bbtheta^p \coloneqq [\bbtheta^{\bbp}; \bbtheta^{\bbgamma}]$ to collect the RRM policy parameters. We can write the Lagrangian then as
%
\begin{align}
   \ccalL_\theta(\bbtheta^p, \tbx, \tblambda) &=  \ccalF(\tbx) - \tblambda^T \ccalG(\tbx; \bbtheta^p).\label{eq_lagrangian_compact}
\end{align}
Observe we have further compacted the objective and constraint functions in $\ccalF(\tbx) \coloneqq {\mathcal{U}}(\bbx) - \frac{\alpha}{2}\| \bbz\|_2^2$ and $\ccalG(\tbx; \bbtheta^p) \coloneqq [\bbx -  \E_{\bbH}[\bbf(\bbH, \bbp(\bbH; \bbtheta^{\bbp}), \bbgamma(\bbH;\bbtheta^{\bbgamma}))]  ;  \bbf_{\min} - \bbz - \bbx]$, respectively.

The optimal parameterized dual value $D_\theta^*$ is given by the solution of the dual problem
%
\begin{equation}\label{eq_p1}
    D_\theta^* = \min_{\tblambda} \max_{\bbtheta^p, \tbx} \left[ \ccalF(\tbx) - \tblambda^T \ccalG(\tbx; \bbtheta^p) \right].
\end{equation}
Given the fact that $\tbp(\bbH; \bbtheta^p)$ defines a subset of policies contained in the unparameterized class of policies in \eqref{eq_slack_problem}, the inner minimization in \eqref{eq_p1} can be upper bounded by
%
\begin{equation}\label{eq_p2}
    D_\theta^* \leq  \min_{\tblambda} \max_{\tbp, \tbx} \left[ \ccalF(\tbx) - \tblambda^T \ccalG(\tbx, \tbp) \right],
\end{equation}
where $\ccalG(\tbx, \tbp) \coloneqq [\bbx -  \E_{\bbH}[\bbf(\bbH, \bbp(\bbH), \bbgamma(\bbH))]  ;  \bbf_{\min} - \bbz - \bbx]$. The term on the right hand side of \eqref{eq_p2} indeed constitutes the unparameterized dual problem in \eqref{eq_dual_problem}. Due to strong duality of the original resilient RRM problem formulation---see \cite[Theorem 1]{ribeiro2012optimal}---we know that $D^* = P^*$ and obtain the upper bound on $D_{\theta}^*$ in \eqref{eq_theorem_param_duality}.

We proceed to derive the lower bound on $D_{\theta}^*$. 
%
%
We add and subtract $\tblambda^T \ccalG(\tbx, \tbp)$ to and from the right hand side of \eqref{eq_p1}:
\begin{align}
    D_\theta^* &= \min_{\tblambda} \max_{\bbtheta^p, \tbx} \Bigg\{ \left[ \ccalF(\tbx) - \tblambda^T \ccalG(\tbx, \tbp)\right] \nonumber \\
    &\qquad\qquad\qquad - \left[ \tblambda^T \left(\ccalG(\tbx; \bbtheta^p) -  \ccalG(\tbx, \tbp)\right) \right] \Bigg\} \label{eq_d3_a} \\
    &= \min_{\tblambda} \max_{\tbx} \Bigg\{ \left[ \ccalF(\tbx) - \tblambda^T \ccalG(\tbx, \tbp)\right] \nonumber \\
    &\qquad\qquad\qquad - \min_{\bbtheta^p} \left[ \tblambda^T \left(\ccalG(\tbx; \bbtheta^p) -  \ccalG(\tbx, \tbp)\right) \right] \Bigg\}, \label{eq_d3_b}
\end{align}
where \eqref{eq_d3_b} is true since the first term on the right hand side of~\eqref{eq_d3_a} does not involve $\bbtheta^p$. Define the term $\Delta_{\theta} \coloneqq  \tblambda^T \left(\ccalG(\tbx; \bbtheta^p) -  \ccalG(\tbx, \tbp)\right)$, we can continue~\eqref{eq_d3_b} as
\begin{align}
    D_\theta^* &= \min_{\tblambda} \max_{\tbx} \left\{ \left[ \ccalF(\tbx) - \tblambda^T \ccalG(\tbx, \tbp)\right] - \min_{\bbtheta^p} \Delta_{\theta} \right\} \\
    &\geq \min_{\tblambda} \max_{\tbx} \left\{ \left[ \ccalF(\tbx) - \tblambda^T \ccalG(\tbx, \tbp)\right] - \min_{\bbtheta^p} |\Delta_{\theta}| \right\}.\label{eq:abs_delta}
\end{align}

%
We proceed to find an upper bound for $|\Delta_{\theta}|$. 
Using H\"{o}lder's inequality, we can write
\begin{align}
    |\Delta_\theta| &\leq  \left\| \tblambda  \right\|_1  \left\|\ccalG(\tbx; \bbtheta^p) -  \ccalG(\tbx, \tbp)\right\|_{\infty}. \label{eq_d5}
\end{align}
A further upper bound can be made from \eqref{eq_d5} by applying Lipschitz continuity of $\ccalG$ in Assumption \ref{assumption_lipschitz} to obtain
\begin{align}
    |\Delta_\theta| &\leq  L_{f} \left\| \tblambda  \right\|_1 \E_{\bbH}\left\|\tbp(\bbH; \bbtheta^p) -  \tbp(\bbH)\right\|_{\infty}, \label{eq_d7}
\end{align}
where $L_f \coloneqq \max\{L_{\bbp}, L_{\bbgamma}\}$. To upper bound the minimum of $|\Delta_{\theta}|$ over $\bbtheta^p$, consider that $\tbp(\cdot; \bbtheta^p)$ is a near-universal parameterization of degree $\eps_p \coloneqq  \max\{ \eps_{\bbp}, \eps_{\bbgamma}\} $. From \eqref{eq_def_bound}, the parameterized primal policy can approximate $\tbp(\bbH)$ to at least a degree of $\eps_p$. From this we obtain
\begin{align}
    \min_{\bbtheta^p} |\Delta_\theta| &\leq \eps_p L_{f} \left\| \tblambda  \right\|_1. \label{eq_d8}
\end{align}
Combining~\eqref{eq_d8} with~\eqref{eq:abs_delta}, we have
\begin{align}
    D_\theta^* &\geq \min_{\tblambda} \max_{\tbx} \left\{ \left[ \ccalF(\tbx) - \tblambda^T \ccalG(\tbx, \tbp)\right] - \eps_p L_{f} \left\| \tblambda  \right\|_1 \right\} \\
    &= \min_{\tblambda} \max_{\tbx}  \left[ \ccalF(\tbx) - \tblambda^T \left(\ccalG(\tbx, \tbp) + \eps_p L_{f} \bbone \right) \right], \label{eq:norm1_lambda}
\end{align}
where in~\eqref{eq:norm1_lambda}, $\bbone$ denotes the vector of all 1's and the equality holds due to the definition of the $\ell_1$-norm and the fact that the dual variables are non-negative. Since~\eqref{eq:norm1_lambda} holds for all $\tbp$, we have
\begin{align}
    D_\theta^* &\geq \min_{\tblambda} \max_{\tbp, \tbx}  \left[ \ccalF(\tbx) - \tblambda^T \left(\ccalG(\tbx, \tbp) + \eps_p L_{f} \bbone \right) \right]. \label{eq:p_back}
\end{align}
Note that the right hand side of~\eqref{eq:p_back} is the dual value of a perturbed version of the problem in~\eqref{eq_slack_problem}, where the constraints are perturbed by $\eps_p L_{f} \bbone$. Since this perturbed problem also has null duality gap, we can use the perturbation inequality in~\cite[\S 5.6.2]{boyd2004convex} to further bound~\eqref{eq:p_back} as
\begin{align}
    D_\theta^* &\geq P^* - \tblambda^{*T} \left(\eps_p L_{f} \bbone \right) \\
    &= P^* - \eps_p L_{f} \left\| \tblambda^*  \right\|_1.
\end{align}
%
This completes the proof. \hfill$\blacksquare$%

%% file: family_configs.tex
\section{Training over a Family of Configurations}\label{appx:family_configs}
As mentioned in Remark~\ref{remark:family_configs}, in practice, we train the RRM policies over a \emph{family} of random network configurations $\bbN \in \ccalN$. Each configuration, drawn from an underlying distribution $\mathfrak{D}_{\bbN}$, represents a random placement of transmitters and receivers and models the long-term channel components stemming from signal attenuation due to the physical distance between the transmitters and receivers, alongside deviations due to obstacles in the environment. We reformulate the resilient RRM formulation in~\eqref{eq_slack_problem} as
\begin{subequations}\label{eq:slack_problem_family_configs}
\begin{alignat}{2}
    &\max_{\bbp,\bbgamma,\bbx,\bbz} &\; & \mathcal \E_{\bbN} \left[ {\mathcal{U}}(\bbx(\bbN)) - \frac{\alpha}{2}\| \bbz(\bbN)\|_2^2 \right],\label{eq:objective_slack_family_configs}             \\
    &~~~~\text{s.t.} && \bbx(\bbN)       \leq  \E_{\bbH} \left[ \bbf(\bbH, \bbp(\bbH), \bbgamma(\bbH)) | \bbN \right],  \ \mathfrak{D}_{\bbN}\text{-a.e.} \label{eq:rate_constraint_family_configs}  \\
    &&& \bbx(\bbN) \geq \bbf_{\min} - \bbz(\bbN), \ \mathfrak{D}_{\bbN}\text{-a.e.}\label{eq:min_rate_constraint_slack_family_configs}\\
    &&& \bbp(\bbH) \in  [0,P_{\max}]^m, \bbgamma(\bbH) \in \Gamma_{n,m}^{\ccalR}, \bbz(\bbN) \geq \bb0\label{eq:resource_constraint_slack_family_configs},%
\end{alignat}
\end{subequations}
where the ergodic average rate $\bbx$ and the slack term $\bbz$ are now \emph{configuration-dependent}. 
Moreover, in~\eqref{eq:min_rate_constraint_slack_family_configs}-\eqref{eq:resource_constraint_slack_family_configs}, $\mathfrak{D}_{\bbN}\text{-a.e.}$ implies that the constraints should be satisfied for almost all large-scale fading configurations $\bbN$ drawn from the distribution $\mathfrak{D}_{\bbN}$. Introducing non-negative dual multiplier functions $\bblambda: \ccalN \rightarrow \reals_+^n$ and $\bbmu: \ccalN \rightarrow \reals_+^n$, we can derive the corresponding Lagrangian function as
\vspace{-.05in}
\begin{align}
   &\ccalL(\bbp,\bbgamma,\bbx,\bbz, \bblambda,\bbmu) \nonumber\\
   &~= \E_{\bbN} \bigg[ {\mathcal{U}}(\bbx(\bbN)) - \frac{\alpha}{2}\| \bbz(\bbN)\|_2^2  \nonumber \\ 
   &~\quad\qquad~-\bblambda(\bbN)^T\left[\bbx(\bbN) - \E_{\bbH} \left[ \bbf(\bbH, \bbp(\bbH), \bbgamma(\bbH)) |\bbN \right] \right]\nonumber \\ 
   &~\quad\qquad~-\bbmu(\bbH^\ell)^T \left[ \bbf_{\min} - \bbz(\bbN) - \bbx(\bbN)\right]\bigg].\label{eq_lagrangian_family_configs}
\end{align}

\vspace{-.05in}
As for the primal-dual learning algorithm, we now draw a set of $B$ network configurations $\{\bbN_{b}\}_{b=1}^{B}$ according to the distribution $\mathfrak{D}_{\bbN}$, and we further draw $B \times T$ fading samples $\{ \bbH_{b,t}\}_{b=1,t=1}^{B,T}$ according to the distribution $\mathfrak{D}_{\bbH}$. For any functions $\ccalF:\ccalN \to \mathbb{R}$ and $\ccalF':\ccalH \to \mathbb{R}$, we define
\vspace{-.05in}
\begin{align}
\hat{\E}_{\bbN} \left[\ccalF(\bbN)\right] &\coloneqq \frac{1}{B} \sum_{b=1}^B \ccalF(\bbN_b),\\
\hat{\E}_{\bbH} \left[\ccalF' (\bbH) | \bbN_b \right] &\coloneqq \frac{1}{T} \sum_{t=1}^T \ccalF'(\bbH_{b,t}).
\end{align}
Then, the primal RRM policy parameters will be updated as
\vspace{-.05in}
\begin{align}
\bbtheta^{\bbp}_{k+1} &= \bbtheta^{\bbp}_k  + \eta_{\bbp} \bbdelta^{\bbp}_k, \label{eq_pd_update1_family_configs}\\
\bbtheta^{\bbgamma}_{k+1} &= \bbtheta^{\bbgamma}_k + \eta_{\bbgamma} \bbdelta^{\bbgamma}_k, \label{eq_pd_update1.5_family_configs}
\end{align}
where $\bbdelta^{\bbp}_k$, and $\bbdelta^{\bbgamma}_k$ are defined as
\vspace{-.05in}
\begin{align*}
\bbdelta^{\bbp}_k &=    \hat{\E}_{\bbN}\bigg[ \nabla_{\bbtheta^{\bbp}} \left\{ \bblambda(\bbN)^T \hat{\E}_{\bbH} \left[ \bbf(\bbH, \bbp(\bbH), \bbgamma(\bbH)) | \bbN \right] \right\}\bigg],  \\
\bbdelta^{\bbgamma}_k &=   \hat{\E}_{\bbN}\bigg[ \nabla_{\bbtheta^{\bbgamma}} \left\{ \bblambda(\bbN)^T \hat{\E}_{\bbH} \left[ \bbf(\bbH, \bbp(\bbH), \bbgamma(\bbH)) | \bbN \right] \right\}\bigg].
\end{align*}
As the remaining primal and dual policies are not parameterized, we update the primal and dual variables corresponding to each training configuration separately. In particular, for any $b\in\{1,\dots,B\}$, let $(\bbx_b, \bbz_b, \bblambda_b, \bbmu_b)$ respectively denote the ergodic average rate, slack and dual variables corresponding to the $b$\textsuperscript{th} configuration, i.e.,
\vspace{-.05in}
\begin{align*}
(\bbx_b, \bbz_b, \bblambda_b, \bbmu_b) = \left(\bbx(\bbN_{b}),\bbz(\bbN_{b}), \bblambda(\bbN_{b}), \bbmu(\bbN_{b})\right).
\end{align*}
Then, for the $b$\textsuperscript{th} configuration, $b\in\{1,\dots,B\}$, we update the ergodic average rate and slack variables as
\vspace{-.05in}
\begin{align}
{\bbx}_{b,k+1} &= {\bbx}_{b,k}  +  \eta_{\bbx} \bbdelta^{\bbx}_{b,k}, \label{eq_pd_update2_family_configs}\\
{\bbz}_{b,k+1} &= \left[{\bbz}_{b,k}  +  \eta_{\bbz} \bbdelta^{\bbz}_{b,k}\right]_+,
\label{eq_pd_update3_family_configs}
\end{align}
where $\bbdelta^{\bbx}_{b,k}$ and $\bbdelta^{\bbz}_{b,k}$ are defined as
\vspace{-.05in}
\begin{align*}
\bbdelta^{\bbx}_{b,k} &=   \nabla_{{\bbx_{b,k}}}\left\{\mathcal{U}(\bbx_{b,k})\right\} + \bbmu_{b,k} - \bblambda_{b,k} .\\
\bbdelta^{\bbz}_{b,k} &=  \bbmu_{b,k}  - \alpha  \bbz_{b,k}.
\end{align*}

Finally, we update the dual variables as
\vspace{-.05in}
\begin{align}
{\bblambda}_{b,k+1} &=  \left[{\bblambda}_{b,k} - \eta_{\bblambda} \bbdelta^{\bblambda}_{b,k}\right]_+,\label{eq_pd_update4_family_configs} \\
{\bbmu}_{b,k+1} &= \left[{\bbmu}_{b,k} - \eta_{\bbmu} \bbdelta^{\bbmu}_{b,k}\right]_+,\label{eq_pd_update5_family_configs}
\end{align}
where $\bbdelta^{\bblambda}_{b,k}$ and $\bbdelta^{\bbmu}_{b,k}$ are defined as
\vspace{-.05in}
\begin{align*}
\bbdelta^{\bblambda}_{b,k} &=    \bbx_{b,k} - \hat{\E}_{\bbH} \left[ \bbf(\bbH, \bbp(\bbH), \bbgamma(\bbH)) | \bbN_b \right], \\
\bbdelta^{\bbmu}_{b,k} &=  \bbf_{\min} - \bbz_{b,k} - \bbx_{b,k}.
\end{align*}